\documentclass[12pt,a4paper]{article}
\usepackage{amssymb,amsmath,amsthm}
\usepackage[dvips]{graphicx,color}

\newcommand{\lcm}{\text{lcm}}
\newcommand{\bartial}{{\bar{\partial}}}

\newcommand{\Natural}{\mathbb{N}}
\newcommand{\Integer}{\mathbb{Z}}
\newcommand{\mat}{\begin{pmatrix}}
\newcommand{\tam}{\end{pmatrix}}

\newcommand{\mc}{\mathcal}
\newcommand{\mf}{\mathfrak}

\newcommand{\WZ}{{\text{WZ}}}

\newcommand{\Ad}{{\text{Ad}}}
\newcommand{\tr}{\text{tr}\,}
\newcommand{\rank}{\text{rank}\,}

\newcommand{\g}{\mathfrak{g}}

\newcommand{\ag}{\mathfrak{\hat{g}}}

\title{\vspace*{1cm}\bf\LARGE Generalised permutation branes\bigskip}
\date{September 2005}
\author{Stefan Fredenhagen$^1$ and Thomas Quella$^2$ \\[10mm]
$^1$ Institut f{\"u}r Theoretische Physik, ETH-H{\"o}nggerberg,\\
CH--8093 Z{\"u}rich, Switzerland\\[2mm]
$^2$ King's College London, Department of Mathematics,\\
Strand, London WC2R 2LS, United Kingdom}

\begin{document}
\begin{titlepage}
\maketitle
\thispagestyle{empty}

\vskip1cm

\begin{abstract} 
We propose a new class of non-factorising D-branes in the product
group $G\times G$ where the fluxes and metrics on the two factors
do not necessarily coincide. They generalise the maximally symmetric
permutation branes which are known to exist when the fluxes agree,
but break the symmetry down to the diagonal current algebra in the
generic case. Evidence for the existence of these branes comes from
a Lagrangian description for the open string world-sheet and from
effective Dirac-Born-Infeld theory. We state the geometry, gauge fields and,
in the case of $SU(2)\times SU(2)$, tensions and partial
results on the open string spectrum. In the latter case the generalised
permutation branes provide a natural and complete explanation
for the charges predicted by K-theory including their torsion.
\end{abstract}

\vspace*{-19cm}\noindent 
{\hfill \tt {KCL-MTH-05-11}} \\
\bigskip\vfill
\noindent

\end{titlepage} 
\setcounter{tocdepth}{2}
\tableofcontents
% -----------------------------------------------------------------------
% -----------------------------------------------------------------------
\section{Introduction}

Group manifolds have widely been used as a playground to study string
theory in non-trivial, curved backgrounds. For compact groups, the
conformal field theory (CFT) on the world-sheet is rational and
completely under control. D-branes in these backgrounds have been
investigated using boundary conformal field theory and semi-classical
methods.
It might surprise that despite all the progress that has been made in
understanding branes on group manifolds, there are usually not
enough D-branes known to explain the whole charge group predicted by
(twisted) K-theory.
\smallskip

  Soon after the essential role of (twisted) K-theory for the
description of decay processes of D-branes was pointed out in
\cite{Minasian:1997mm,Witten:1998cd,Bouwknegt:2000qt} there have been
attempts to explain the K-groups for group manifolds
using boundary renormalisation group flow ideas. In this way the K-groups
for $SU(2)$ and $SU(3)$ could be understood completely~\cite{Alekseev:2000jx,Fredenhagen:2000ei,Gaberdiel:2003kv}. In the case of
$SU(2)$ the only type of charge is associated with Cardy branes
\cite{Cardy:1989ir} while for $SU(3)$ there also exists a second type
of charge which could be assigned to twisted branes
\cite{Birke:1999ik}. Both classes of branes are maximally
symmetric. Later the K-groups have been calculated in a purely
mathematical manner \cite{Braun:2003rd,Freed:2003qx,Douglas:2004}.
According to these results the number of charges grows exponentially
with the rank of the (simple) group $G$, but this exceeds the
different types of maximally symmetric branes by far.  
Although there are some ideas what the branes carrying the remaining
charges could be (see~\cite{Gaberdiel:2004hs,Gaberdiel:2004za} for a
proposal for $SU (n)$ based on the symmetry breaking branes
of~\cite{Maldacena:2001ky,Quella:2002ct,Quella:2002ns}), it is fair to
say that a satisfactory answer is still missing.
\smallskip

A simple example where one can hope to resolve the mismatch and
explain the complete charge group is the Wess-Zumino-Novikov-Witten (WZNW)
model on the product
group $SU(2)\times SU(2)$. This theory has two parameters, the levels
$k_1$ and $k_2$, which control the size and the H-field of the
background. The corresponding K-theory is twisted by the third
cohomology class $\tau =k_{1}\tau_{1}+k_{2}\tau_{2}$ where $\tau_{1}$
and $\tau_{2}$ are the generators of the third integer cohomology
group of the two factors.\footnote{Here, levels are
understood as levels of the supersymmetric model.} It is given by (see
appendix~\ref{app:Ktheory})
\begin{equation}\label{KtheorySU2}
^{\tau }K\bigl(SU (2)\times SU (2)\bigr) \ =\ \mathbb{Z}_{k}
\oplus \mathbb{Z}_{k} \ \ , 
\end{equation}
where $k=\gcd (k_{1},k_{2})$ is the greatest common divisor of the two
levels. One summand in~\eqref{KtheorySU2} belongs to $0$-branes and
can readily be explained by
factorising Cardy branes; the charge relations derived from boundary
renormalisation group flows as
in~\cite{Alekseev:2000jx,Fredenhagen:2000ei} produce the correct order
$k$ of the charge group. The other summand corresponds to a three-dimensional cycle with
vanishing integrated H-flux. When the two levels are equal, this cycle
is wrapped by permutation branes~\cite{Recknagel:2002qq} (see also
\cite{Figueroa-O'Farrill:2000ei,Gaberdiel:2002jr,Sarkissian:2003yw} for
more specific examples). The existence of the
latter relies on an order-2 automorphism of the affine Lie algebra
$\widehat{su}(2)_{k}\oplus \widehat{su} (2)_{k}$ which
interchanges the two factors. If the levels are different, no such
automorphism exists, and there is no obvious CFT construction of
D-branes corresponding to the second summand in the K-group. Let us
nevertheless emphasise that we are in a special situation where the
mismatch between K-theory and the known D-branes is cured for a
particular choice of parameters ($k_{1}=k_{2}$) and one can hope that
this coincidence provides some hints regarding the appropriate
construction of the missing branes for an arbitrary choice of the
levels.  
\smallskip

Following this reasoning we give a proposal for the branes that
explain the missing charges. Instead of using the established
constructions for symmetry breaking branes\footnote{One could for
instance break the $SU (2)$ symmetry down to the coset $SU (2)/U (1)$
and $U(1)$ and then perform a twist in the $U(1)$ part, but the corresponding
branes seem to have the wrong topology to explain the charges.}, we are
directly led by the geometry of the branes that we expect from
K-theory. We suggest the existence of ``generalised permutation
branes'' in products $G\times G$ of Lie groups with different
levels. The geometry of the simplest new brane is given
by
\begin{equation}
  \bigl\{\,(\,g^{k_2/k},g^{-k_1/k}\,)\,\bigr|\,g\in G\,\bigr\}
  \ \subset\ G\times G \ \ .
\end{equation}
It represents a $(\dim G)$-dimensional submanifold with vanishing
total H-flux and reduces to the geometry of an ordinary permutation brane
if $k_1=k_2$. This brane has higher-dimensional cousins which are
stabilised by a non-trivial F-field. They are described in
section~\ref{sc:Genpermbranes}.
\smallskip

To support our proposal we derive the world-sheet Lagrangian
and employ semi-classical target space methods
(Dirac-Born-Infeld). The branes are not maximally symmetric, but the
symmetry of the diagonally embedded group $G$ is still conserved. For
given levels, there are only finitely many generalised permutation
branes. Their number coincides with the number of untwisted maximally
symmetric branes in a single group factor $G$ with level $\kappa=\lcm
(k_{1},k_{2})$, the least common multiple of $k_{1}$ and $k_{2}$. In
the case of $SU(2)\times SU(2)$ we also analyse tensions and charges of the new
branes. The relative tensions are controlled by the modular S-matrix
of a single $SU (2)$-model with level $\kappa$.  Stability
considerations indicate that the charge of the branes is
conserved modulo $k=\gcd (k_{1},k_{2})$ which is in perfect agreement
with the K-theory result.  
\medskip

The plan of the paper is as follows. In section 2 we discuss the 
$\sigma$-model on a world-sheet with boundaries. After a short
introduction into WZNW models, we formulate the geometry
of our proposed generalised permutation branes in the product group
$G\times G$ and give an expression for the boundary
two-form living on them. Then the quantisation condition which leads to a
discrete brane spectrum is discussed and it is shown that the branes
preserve the diagonal $G$-symmetry.
\smallskip

Section 3 concentrates on the case of $SU (2)\times SU (2)$. First we
analyse the geometry of the branes in detail and show that only a subset
of the world volumes avoid self-intersections and can be considered
stable. The instability of the other branes reproduces the expected charge
group including its torsion. In the following we discuss the effective Dirac-Born-Infeld
description. It is shown that the proposed branes satisfy the equations of
motions, partly using numerical methods. Also the tensions and the open
string spectrum are analysed numerically.
\smallskip

Section 4 generalises some of the results of section 3 to higher rank
groups. A summary and a discussion of open problems and future
directions conclude the paper. In particular, we comment on a possible extension of
our results to cosets where generalised permutation branes recently appeared
in the rather complementary approach of matrix factorisations in the
topological subsector of products of $N=2$ minimal models
\cite{Brunner:2005fv}. Two appendices contain the computation of the
relevant K-groups and the technical details of the DBI calculations.

% -----------------------------------------------------------------------
% -----------------------------------------------------------------------
% -----------------------------------------------------------------------
\section{Lagrangian description}

% -----------------------------------------------------------------------
% -----------------------------------------------------------------------
\subsection{Wess-Zumino-Novikov-Witten models}\label{sc:WZW}

The propagation of strings on a group manifold $\mc{G}$ is described
by a WZNW model \cite{Witten:1984ar,Gepner:1986wi}, a special and conformally
invariant instance of
non-linear $\sigma$-models. Since we are interested in open strings,
the world-sheet $\Sigma$ has a non-trivial boundary. For our purposes it is
sufficient to assume that $\partial\Sigma$ consists of precisely one component
with the topology of a circle. At this boundary, the group-valued field
$g$ on $\Sigma$ is constrained to some subset $\mc{D}\subset \mc{G}$, the
D-brane. The action reads \cite{Alekseev:1998mc,Gawedzki:1999bq} 
\begin{equation}
  \label{eq:WZNW}
  \mc{S}^{\mc{G}}[g]
  \ =\ \frac{1}{4\pi i}\int_\Sigma\bigl\langle g^{-1}\partial g,g^{-1}\bartial g\bigr\rangle\,dz\wedge d\bar{z}
       +\frac{1}{4\pi i}\int_B\omega^{\text{WZ}}
       -\frac{1}{4\pi i}\int_D\omega_C\ \ .
\end{equation}
Here, $\langle\cdot,\cdot\rangle$ is a non-degenerate invariant
bilinear form on the associated Lie algebra, and the two-dimensional
disc $D$ is used to fill the hole in $\Sigma$, such that $\Sigma \cup D$
has no boundary. As usual the Wess-Zumino term is
defined as an integral of the Wess-Zumino form
\begin{align}
  \omega^{\text{WZ}}(g)\ =\ \frac{1}{6}\,\bigl\langle g^{-1}dg,[g^{-1}dg,g^{-1}dg]\bigr\rangle
\end{align}
  over a three-dimensional space $B$ with $\partial B=\Sigma\cup D$.
  The theory should not depend on the choice of the two auxiliary
  manifolds $D$ and $B$ and the continuation of the field $g$ to $D$ and
  $B$. This requires the Wess-Zumino form to be integral and trivialised
  by the boundary two-form on the brane $\mc{D}$,
\begin{equation}
  \omega^{\text{WZ}}\bigr|_{\mc{D}}\ =\ d\omega_C\ \ .
\end{equation}
  The discussion of a further quantisation condition for $\omega_C$
  will be postponed until section \ref{sc:Quantisation}.  
  If all these constraints are satisfied, the action \eqref{eq:WZNW}
  is well-defined up to multiples of $2\pi i$ which do not affect the
  path integral. We refer to the literature for details.
\smallskip

  In this paper we are interested in string theory on the product
group $\mc{G}=G\times G$ where $G$ is a compact simply-connected
simple group. If we denote by `$\tr$' the suitably
normalised Killing form on $\g=\text{Lie}(G)$, then the most general non-degenerate
invariant bilinear form on the product is given by
\begin{equation}
  \label{eq:Form}
  \bigl\langle(X_1,X_2),(Y_1,Y_2)\bigr\rangle
  \ =\ k_1\,\tr(X_1Y_1)+k_2\,\tr(X_2Y_2)
\end{equation}
  with non-vanishing real numbers $k_i$. The integrality of the Wess-Zumino
  form and unitarity of the (Minkowskian version of the) theory imposes
  the constraint that the levels $k_i$ are non-negative integers.

% -----------------------------------------------------------------------
% -----------------------------------------------------------------------
\subsection{\label{sc:Genpermbranes}Generalised permutation branes}

  In each group $\mc{G}$ there exist so-called maximally symmetric branes
  which are associated to an automorphism $\Omega$ of $\mc{G}$ provided the
  latter may be lifted to the underlying affine Lie algebra of the WZNW model.
  On the CFT side these branes have been constructed in
  \cite{Cardy:1989ir,Birke:1999ik}. On the geometric side they have later
  been shown to wrap twisted conjugacy classes
  \cite{Alekseev:1998mc,Felder:1999ka}
\begin{equation}
  \mc{C}_\Omega(f)\ =\ \bigl\{gf\Omega(g^{-1})\,\bigl|\,g\in\mc{G}\bigr\}\ \ .
\end{equation}
  These submanifolds admit an obvious action of $\mc{G}$ which corresponds
  to the affine symmetry preserved in the boundary CFT.
\smallskip

  Let us now focus on the product group $\mc{G}=G\times G$. It can easily be
  understood that there are maximally symmetric branes which completely
  factorise. They correspond to automorphisms which may be written as
  a product of automorphisms. On the other hand the group $\mc{G}$ also
  admits the permutation automorphism
\begin{equation}
  \label{eq:Twist}
  \tau(g_1,g_2)\ =\ (g_2,g_1)\ \ .
\end{equation}
  The associated twisted conjugacy classes have the form
\begin{equation}
  \label{eq:TwConjEq}
  \mc{C}_\tau(f)\ =\
\bigl\{(h_1fh_2^{-1},h_2fh_1^{-1})\,\bigl|\,h_1,h_2\in G\bigr\} \ \ , 
\end{equation}
  and one might believe that they are good candidates for branes.
  This assertion is certainly true for $k_1=k_2$. Whenever the
  two levels disagree, however, the permutation automorphism $\tau$ does not leave the
  metric invariant and cannot be lifted to an automorphism of the affine Lie algebra
  $\ag_{k_1}\oplus\ag_{k_2}$ which underlies the WZNW model. We thus
  conclude that the twisted conjugacy classes \eqref{eq:TwConjEq}
  just describe the loci of branes if $k_1=k_2$.\footnote{Reading \cite{Bordalo:2001ec}
  one might get the impression that {\em all} twisted conjugacy classes are loci
  of D-branes. In their analysis, however, the authors {\em implicitly} assumed
  that the automorphism preserves the scalar product. For simple groups
  all automorphisms possess this property.}
\smallskip

  The main result of our paper is that a slight modification of the
  geometry \eqref{eq:TwConjEq} actually gives rise to proper
  D-branes even for $k_1\neq k_2$. To be more concrete,
  we propose that the submanifold\footnote{This proposal came up in discussions with A.\ Alekseev.}
\begin{equation}
  \label{eq:TwConjNeq}
  \mc{D}_\tau(f)
  \ =\ \bigl\{\bigl((h_1fh_2^{-1})^{k'_2},(h_2fh_1^{-1})^{k'_1}\bigr)\,\bigl|\,h_1,h_2\in G\bigr\}\ \ ,
\end{equation}
  where $k'_i=k_i/k$ and $k=\gcd(k_1,k_2)$, is the locus of a
  {\em generalised permutation brane} if we impose suitable
  quantisation conditions on the constant $f$ which may be
  chosen from a fixed maximal torus $T\subset G$. Obviously, this
  geometry reduces to the twisted conjugacy class \eqref{eq:TwConjEq}
  whenever the two levels coincide. In general
  the dimension of the branes is $(2\dim G-\rank G)$ but
  for certain degenerate values of $f$ it will have a lower
  value. In particular if $f$ equals the identity element $e\in G$,
  the expression \eqref{eq:TwConjNeq} simplifies to
\begin{equation}
  \label{eq:TwConjNeqId}
  \mc{D}_\tau(e)
  \ =\ \bigl\{\bigl(g^{k'_2},g^{-k'_1}\bigr)\,\bigl|\,g\in G\bigr\}\ \ .
\end{equation}
  In this case the dimension is given by $\dim G$. Since the constants
  $k'_i$ are relatively prime, there is a one-to-one correspondence
  between elements of $\mc{D}_\tau(e)$ and elements of $G$.  
\smallskip

  The reason for the occurrence of the non-trivial exponents
  $k'_1$ and $k'_2$ in the definition \eqref{eq:TwConjNeq} becomes
  immediately obvious if we remember that we have to find a boundary
  two-form $\omega_C$ which trivialises the Wess-Zumino form
  on the brane. Repeatedly applying the Polyakov-Wiegmann identity
\begin{equation}
  \label{eq:PW}
  \omega^{\text{WZ}}(hg)
  \ =\ \omega^{\text{WZ}}(h)+\omega^{\text{WZ}}(g)
       -d\,\tr\bigl(h^{-1}dh\wedge dgg^{-1}\bigr)
\end{equation}
  one can easily prove
\begin{equation}
  \omega^{\text{WZ}}(g^n)
  \ =\ n\,\omega^{\text{WZ}}(g)
+\sum_{j=1}^{n-1}(n-j)\,d\,\tr\bigl(\Ad_g^j(g^{-1}dg)\wedge g^{-1}dg\bigr)\ \ . 
\end{equation}
  Here, $\Ad_{g} (X)=gXg^{-1}$ denotes the adjoint action of
  the group element $g$ on $X\in\mf{g}$. 
  If we instead consider the total Wess-Zumino form
  $\omega^{\text{WZ}}(g_1,g_2)=k_1\omega^{\text{WZ}}(g_1)
  +k_2\omega^{\text{WZ}}(g_2)$ and restrict it to the submanifold
  \eqref{eq:TwConjNeq}, one realises that the constants $k'_i$
  have been chosen in a way such that the terms which cannot be
  written as a total derivative cancel out.
  As a consequence we find that
\begin{equation}
  \omega^{\text{WZ}}\bigr|_{\mc{D}_\tau(f)}=\ d\omega_C
\end{equation}
  is satisfied if we choose the boundary two-form $\omega_C$
  according to
\begin{equation}
  \label{eq:OmegaGen}
  \begin{split}
    \omega_C
    &\ =\ \frac{k_1k_2}{k}\,\Bigl\{\tr\bigl(h_1^{-1}dh_1\wedge\Ad_f(h_2^{-1}dh_2)\bigr)+\tr\bigl(h_2^{-1}dh_2\wedge\Ad_f(h_1^{-1}dh_1)\bigr)\Bigr\}\\[2mm]
    &\qquad\quad+k_1\sum_{j=1}^{k'_2-1}(k'_2-j)\,\tr\bigl(\Ad_{g^j}(g^{-1}dg)\wedge g^{-1}dg\bigr)_{g=h_1fh_2^{-1}}\\[2mm]
    &\qquad\quad+k_2\sum_{j=1}^{k'_1-1}(k'_1-j)\,\tr\bigl(\Ad_{g^j}(g^{-1}dg)\wedge g^{-1}dg\bigr)_{g=h_2fh_1^{-1}}\ \ .
  \end{split}
\end{equation}
  Note that the first two terms compensate each other if $f$ is central.
  If in addition the two levels coincide, the boundary two-form vanishes
  identically.
\smallskip

  Let us finally comment on the symmetries of our generalised
  permutation branes. We already mentioned that the branes are maximally
  symmetric in the case of equal levels $k_1=k_2=k$, i.e.\ they
  preserve the current algebra $\ag_k\oplus\ag_k$. On the geometric
  side this corresponds to the invariance of the brane world volume
  \eqref{eq:TwConjEq} under the two different twisted adjoint actions
\begin{align}
  (g_1,g_2)&\ \mapsto\ (hg_1,g_2h^{-1})&
  &\text{ and }&
  (g_1,g_2)&\ \mapsto\ (g_1h^{-1},hg_2)\ \ .
\end{align}
  For $k_1\neq k_2$, however, the world volumes \eqref{eq:TwConjNeq}
  and \eqref{eq:TwConjNeqId} are not invariant anymore under this
  action of $G\times G$ because of the non-trivial exponents $k'_i$.
  Instead they only admit an action of the diagonal subgroup
\begin{equation}
  (g_1,g_2)\ \mapsto (hg_1h^{-1},hg_2h^{-1})
\end{equation}
  in that case. We thus
  expect that the affine symmetry is at least broken down to
  $\ag_{k_1+k_2}$. In fact we will argue in section \ref{sc:Gluing} that
  this smaller symmetry is indeed conserved.\footnote{Note that in
  the CFT description
  the branes have to preserve the Virasoro algebra associated with
  $\ag_{k_1}\oplus\ag_{k_2}$. Since the central charge of the diagonal
  current algebra $\ag_{k_1+k_2}$ is not sufficient,
  there also has to be an additional symmetry including a Virasoro
  algebra with the remaining central charge. So far we could not identify
  this additional symmetry but according to our geometric arguments it will
  quite certainly not be of affine type.}
\smallskip

  We conclude that generalised permutation branes provide a new, geometrically
  motivated, class of
  symmetry breaking D-branes. In particular we wish to emphasise that
  the non-maximally symmetric branes constructed in
  \cite{Quella:2002ct,Quella:2002ns} have a geometrical interpretation 
  which is very distinct from \eqref{eq:TwConjNeq}. In addition, those
  branes also exist for $k_1=k_2$ and certainly do not coincide with
  ordinary permutation branes in that case.

% -----------------------------------------------------------------------
% -----------------------------------------------------------------------
\subsection{\label{sc:Quantisation}Brane quantisation}

In the previous section we described the geometry of generalised
permutation branes. The branes are labelled by an element $f$
of the Cartan torus of $G$. As always for branes in WZNW models, 
there is a quantisation condition which restricts the
possible values of $f$ to a discrete set. It arises from an
ambiguity in the Lagrangian description whenever the brane geometry
has a non-trivial two-cycle (see
e.g.\ \cite{Klimcik:1996hp,Alekseev:1998mc,Gawedzki:1999bq}).
Namely, the action $\mc{S}^{\mc{G}}$ in~\eqref{eq:WZNW} depends on how the map
$g$ is continued from $\Sigma$ to the whole
$\Sigma\cup D $. The ambiguity takes the form
\begin{equation}
  \Delta \mc{S}^{\mc{G}} \ =\ \frac{1}{4\pi i}\left( \int_{D_{3}} \omega^{\WZ} -
  \int_{S^{2}} \omega_{C} \right) 
\end{equation}
for a three dimensional ball
$D_{3}$, whose boundary $S^{2}$ is mapped into the brane. To make the
path integral well defined, $\Delta \mc{S}^{\mc{G}}$ has to be a multiple of
$2\pi i$. In other words, the F-field $F=\omega_{C}-B$ (where $B$ is a
two-form such that $dB=\omega^{\WZ}$) must have quantised flux through any
two-sphere embedded into the brane~\cite{Bachas:2000ik}. 
\smallskip

In~\cite{Gawedzki:1999bq,Bordalo:2001ec} the quantisation condition
has been worked out for untwisted branes in arbitrary compact simple
simply-connected Lie groups~$G$. The allowed branes are localised on
conjugacy classes $\mc{C}(t)\subset G$ of elements
\begin{equation}\label{quantisedt}
  t\ =\ \exp \frac{2\pi i\lambda}{k}
\end{equation}
in the Cartan torus where $\lambda$ is an integral
weight of $\mf g$.
\smallskip

The generalised permutation branes $\mc{D}_{\tau}(f)$ in the product
group are described by an immersion of $G\times \mc{C}(f^2)$ into
$G\times G$,
\begin{equation}\label{higherbranesmap}
  G\times \mc{C}(f^2) \ni
(g,c)\,\mapsto\,\bigl(g^{k'_{2}},(cg^{-1})^{k'_{1}}\bigr)\in 
\mc{D}_{\tau } (f) \ \ .
\end{equation}
Ignoring possible self-intersections (see
section~\ref{sc:intersections}), the only non-trivial two-cycles in
the brane come from the conjugacy class $\mc{C}(f^2)$. So we can
restrict our analysis to the submanifold $\{ e\}\times \mc{C}(f^2)$
which is embedded in $G\times G$ as $\{e\}\times
(\mc{C}(f^2))^{k'_{1}} = \{e\} \times \mc{C}(f^{2k'_1})$. On
this conjugacy class the boundary two-form $\omega_{C}$ restricts to
the usual two-form of untwisted branes in a single copy of $G$ with
level $k_{2}$. Employing the result~\eqref{quantisedt}, we find that
$t=f^{2k'_{1}}$ can take the quantised values $\exp \frac{2\pi
i\lambda}{k_{2}}$, i.e.\ $f$ is restricted to
\begin{equation}
f\ =\ \exp \frac{\pi i\lambda}{\kappa}
\end{equation}
where $\kappa=kk'_{1}k'_{2}$ is the least common multiple of the
levels $k_{1}$ and $k_{2}$.
\smallskip

The number of generalised permutation branes thus coincides with the number
of untwisted branes in a single copy of $G$ with level $\kappa$.
Surprisingly enough this suggests that the exact CFT description will
be based on a {\em rational} conformal subalgebra of the bulk symmetry
$\ag_{k_1}\oplus\ag_{k_2}$ and gives a severe restriction on the complete
set of symmetries of the brane.

% -----------------------------------------------------------------------
% -----------------------------------------------------------------------
\subsection{\label{sc:Gluing}Discussion of gluing conditions}

To better understand the new branes, we would like to inspect the
gluing conditions on the boundary. Let us assume that $\Sigma$ is the
upper half-plane. The conserved currents of the bulk-model are
\begin{alignat*}{2}
J_{j} (z)\ &=\ -k_{j}\partial g_{j} g_{j}^{-1} \ &&=\
-\tfrac{k_{j}}{2}\partial_{x}g_{j} g_{j}^{-1}  +\tfrac{ik_{j}}{2}
\partial_{y}g_{j} g_{j}^{-1} \\
\intertext{and}
\bar{J}_{j} (z)\ &=\ k_{j}g_{j}^{-1 }\bar{\partial} g_{j} \ &&=\
\tfrac{k_{j}}{2}g_{j}^{-1}\partial_{x}g_{j}  +\tfrac{ik_{j}}{2}
g_{j}^{-1}\partial_{y}g_{j} 
 \ \ ,
\end{alignat*}
where $g_1(z,\bar{z})$ and $g_2(z,\bar{z})$ are the bulk fields. 
At the boundary, the currents $J_{j}$ and $\bar{J}_{j}$ are related.
We will show in the following that the diagonal combination $J_{1}+J_{2}$
equals the anti-holomorphic counterpart $\bar{J}_{1}+\bar{J}_{2}$ at
$z=\bar{z}$. 

From the brane geometry~\eqref{eq:TwConjNeq}, we find a condition on
$\partial_{x}g_{i}$: for $y=0$ we have $g_{1}=
(h_{1}fh_{2}^{-1})^{k_{2}'}$ and $g_{2}= (h_{2}fh_{1}^{-1})^{k_{1}'}$,
hence
\begin{equation}
  \label{geomrel}
  \begin{split}
    g_{1}^{-1}\partial_{x}g_{1}
    &\ =\ \frac{1-\Ad_{g_{1}}^{-1}}{1-\Ad_{h_{1}fh_{2}^{-1}}^{-1}} 
          \bigl((h_{1}fh_{2}^{-1})^{-1}\partial_{x} (h_{1}fh_{2}^{-1})\bigr)\\[2mm]
    g_{2}^{-1}\partial_{x}g_{2}
    &\ =\ \frac{1-\Ad_{g_{2}}^{-1}}{1-\Ad_{h_{2}fh_{1}^{-1}}^{-1}}
          \bigl((h_{2}fh_{1}^{-1})^{-1}\partial_{x} (h_{2}fh_{1}^{-1})\bigr) \ \ .
  \end{split}
\end{equation}
This provides a relation between $g_{1}^{-1}\partial_{x}g_{1}$ and
$g_{2}^{-1}\partial_{x}g_{2}$ at the boundary.
\smallskip

For a further relation we have to study the equations of motion. We
vary the action and restrict our attention to the contribution from the
boundary. From the kinetic term we find
\begin{equation}
\delta \mc{S}_{\text{kin}}^{\mc{G}}\bigr|_{\text{boundary}}\ =\ \frac{k_{1}}{4\pi}\int
dx\ \tr g_{1}^{-1}\delta g_{1}\, g_{1}^{-1}\partial_{y}g_{1} \, +\, 
\frac{k_{2}}{4\pi}\int
dx\ \tr g_{2}^{-1}\delta g_{2}\, g_{2}^{-1}\partial_{y}g_{2}\ \ .
\end{equation}
The variations of $g_{1}$ and $g_{2}$ on the boundary are related,
because $g_{1}= (h_{1}fh_{2}^{-1})^{k_{2}'}$ and $g_{2}=
(h_{2}fh_{1}^{-1})^{k_{1}'}$. For our purpose we do not need the full
boundary equations of motion, so we concentrate on variations such that 
\begin{equation}
\delta h_{1}\,h_{1}^{-1}\ =\ \delta h_{2}\,h_{2}^{-1}\ =\ \delta h \,h^{-1}\ \ .
\end{equation}
  Under this assumption we obtain
\begin{equation}
\delta \mc{S}_{\text{kin}}^{\mc{G}}\bigr|_{\text{boundary}}\ =\ -\frac{1}{4\pi}\int dx \
\tr (\delta h h^{-1})\bigl[k_{1} (1-\Ad_{g_{1}})
(g_{1}^{-1}\partial_{y}g)+k_{2}  (1-\Ad_{g_{2}})
(g_{2}^{-1}\partial_{y}g_{2}) \bigr]\ \ .
\end{equation}
There is another contribution to $\delta S$ of that form which comes
from the integral over the boundary two-form $\omega_{C}$ given
in~\eqref{eq:OmegaGen}. We find
\begin{multline*}
\delta \mc{S}_{\text{WZ}}^{\mc{G}}\bigr|_{\text{boundary}}\ =\ \frac{i}{4\pi }
\int dx\ \tr (\delta h h^{-1})\Big\{-k_{1}k'_{2} \big[ \Ad_{h_{1}f}
(h_{2}^{-1}\partial_{x}h_{2})-\Ad_{h_{2}f^{-1}} 
(h_{1}^{-1}\partial_{x}h_{1})\big]\\
+\big( 1-\Ad_{g}\big)
\sum_{j=1}^{k_{2}'-1}k_{1}
(k_{2}'-j)\Big(\Ad_{g}^{-j} -\Ad_{g}^{j}\Big)
\big( g^{-1}\partial_{x}g \big) 
\Big\}_{g=h_{1}fh_{2}^{-1}} \ +\ (1\leftrightarrow 2)\ \ .
\end{multline*}
This leads to the equation of motion
\begin{align*}
&k_{1} (1-\Ad_{g_{1}}) (g_{1}^{-1}\partial_{y}g)
+k_{2} (1-\Ad_{g_{2}}) (g_{2}^{-1}\partial_{y}g_{2})\\
&\quad  =\ 
\bigg\{-ik_{1}k'_{2}\big[\Ad_{h_{1}f} (h_{2}^{-1}\partial_{x}h_{2})
-\Ad_{h_{2}f^{-1}} (h_{1}^{-1}\partial_{x}h_{1}) \big]
-ik_{1}k'_{2} (1+\Ad_{g})
\big(g^{-1}\partial_{x}g  \big)\\
& \quad \quad \quad \quad 
+ik_{1}(1+\Ad_{g_{1}})\frac{1-\Ad_{g_{1}}^{-1}}{1-\Ad_{g}}
\big(g^{-1}\partial_{x} g \big)
\bigg\}_{g=h_{1}fh_{2}^{-1}} \ +\  (1\leftrightarrow 2)\ \ .
\end{align*}
Using the relations~\eqref{geomrel} from the brane geometry we obtain
\begin{equation}
k_{1} (1-\Ad_{g_{1}})(g_{1}^{-1}\partial_{y}g_{1})
+ (1\leftrightarrow 2)\    
=\ ik_{1} (1+\Ad_{g_{1}}) (g_{1}^{-1}\partial_{x}g_{1}) 
+ (1\leftrightarrow 2)\ \ .
\end{equation}
Expressing the result in terms of the currents $J_{j}$ and
$\bar{J}_{j}$ we finally arrive at the desired boundary
condition
\begin{equation}\label{diaggluing}
J_{1}+J_{2}\ =\ \bar{J}_{1}+\bar{J}_{2}\ \ .
\end{equation}
We expect that the associated symmetry of the classical boundary
theory continues to hold in the full quantum theory.

% -----------------------------------------------------------------------
% -----------------------------------------------------------------------
% -----------------------------------------------------------------------
\section{Generalised permutation branes in $\boldsymbol{SU (2)\times SU (2)}$}

  In this section we will analyse some of the properties of our
generalised permutation branes \eqref{eq:TwConjNeq}. For technical
reasons we restrict ourselves to the simplest and physically most
interesting case, the product group $SU(2)_{k_1}\times
SU(2)_{k_2}$. First we present a detailed discussion of the geometry
of the branes. We argue that the occurrence of self-intersections for
certain values of the brane labels causes instabilities and allows us
to predict the torsion of the brane charges. Afterwards we provide
additional support for the existence of the new branes using the
Dirac-Born-Infeld approach. Finally, the tensions of the
branes and their low-energy excitations are discussed.

% -----------------------------------------------------------------------
% -----------------------------------------------------------------------
\subsection{\label{sc:intersections}Self-intersections and charges}

Let us take a closer look at the proposed brane
geometry. For the simplest brane, the map
\begin{equation}
  G\,\to\,G \times G: \quad  g\,\mapsto\,\bigl(g^{k'_{2}},g^{-k'_{1}}\bigr)
\end{equation}
provides a smooth embedding of $G$ into $G\times G$. Note that the map
is one-to-one: from $g_{1}=g^{k'_{2}}$ and $g_{2}=g^{-k'_{1}}$ one can
reproduce $g$ by taking the product of appropriate powers of $g_{1}$
and $g_{2}$ ($k'_{1}$ and $k'_{2}$ are coprime). For the higher
dimensional branes, we have a parametrisation by the
map~\eqref{higherbranesmap} from $G\times \mc{C}(f^2)$ into $G\times
G$ where $\mc{C}(f^2)\subset G$ is the conjugacy class of $f^{2}$ in
$G$.  As long as the element $f$ of the Cartan torus is close to the
group unit, the situation is very similar to the case of the simplest
brane, and the map is still one-to-one. For larger conjugacy classes
$\mc{C}(f^2)$, however, the map ceases to be one-to-one and the brane
develops self-intersections. This signals an instability of the
brane. In the following we will take $SU (2)\times SU (2)$ as an
example and work out the precise conditions under which intersections
occur and what the instabilities imply for the torsion of the brane
charges.  \smallskip

\begin{figure}
\begin{center}
\includegraphics[width=8cm]{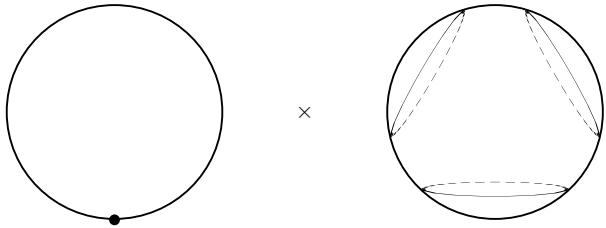}
\end{center}
\caption{\label{fig:genpermb}The intersection of
$\mc{D}_\tau^{\text{red}}(f)$ with the slice $\{e \}\times SU (2)$ for
``reduced'' levels $k'_{1}=1$ and $k'_{2}=3$: The group manifold of $SU (2)$ which
is a three-sphere $S^{3}$ is drawn as a two-sphere, the circles
represent spherical conjugacy classes.}
\end{figure}

The brane geometry is invariant under the adjoint action of the
diagonally embedded $SU(2)$. This action cannot produce any
self-intersections, so we can restrict our discussion to a generating
set of the full orbit which can be chosen to be 
\begin{equation}
  \mc{D}_\tau^{\text{red}}(f)
  \ =\ \bigl\{\bigl(t^{k'_{2}},(ct^{-1})^{k'_{1}}\bigr)
       \,\bigr|\,t\in T,\ c\in \mc{C}(f^2) \bigr\} \ \ .
\end{equation}
Let us for a moment restrict to a simple case and assume that
$k'_{1}=1$. To get an idea of the geometry we look at the slice $\{e
\}\times SU(2)\subset SU(2)\times SU(2)$. What is the intersection of
$\mc{D}_\tau^{\text{red}}(f)$ with this slice? The equation $t^{k'_{2}}=e$
has $k'_{2}$ solutions, namely 
\begin{equation}
t_{j}\ =\ \mat
e^{\frac{2\pi ij}{k'_{2}}} & 0\\
0 & e^{-\frac{2\pi ij}{k'_{2}}}
\tam \quad \quad j=0,\dots , k'_{2}-1 \ \ .
\end{equation}
The restriction of $\mc{D}_\tau^{\text{red}}(f)$ to the slice is the
superposition of $k'_{2}$ copies of the conjugacy class $\mc{C}(f^2)$
where each copy is translated by $t_{j}^{-1}$.
In figure~\ref{fig:genpermb} the resulting geometry is illustrated
for $k'_{2}=3$.
\smallskip

\begin{figure}
\begin{center}
\includegraphics[width=8cm]{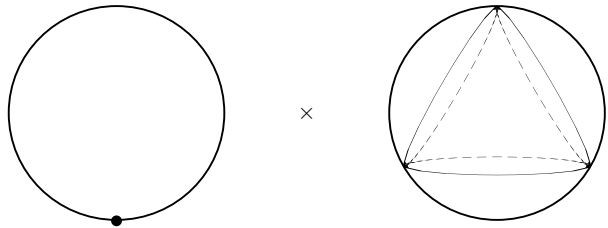}
\end{center}
\caption{\label{fig:genpermc}The brane geometry at the critical angle.}
\end{figure}

\begin{figure}
\begin{center}
\includegraphics[width=8cm]{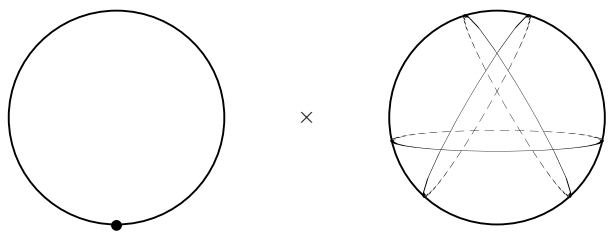}
\end{center}
\caption{\label{fig:genperma}The brane geometry above the critical angle.}
\end{figure}

We parametrise $f$ by
\begin{equation}
f^{2}\ =\ \mat
e^{i\xi}& 0 \\
0 & e^{-i\xi}
\tam \ \ .
\end{equation}
One can immediately see that self-intersections will occur when the
angle $\xi$ reaches the critical value
$\xi^{c}=\frac{\pi}{k'_{2}}$. Figures~\ref{fig:genpermc}
and~\ref{fig:genperma} show the geometry for $\xi$ at and above the
critical angle.
\smallskip

\begin{figure}
\begin{center}
\includegraphics[width=8cm]{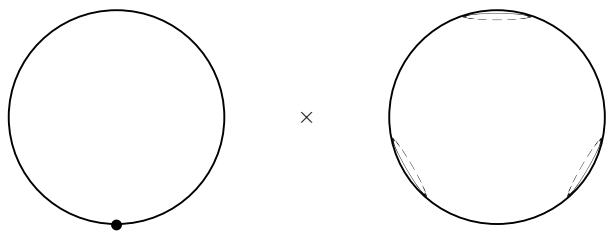}
\end{center}
\caption{\label{fig:genpermd}The brane of figure~\ref{fig:genperma}
might decay into this brane of lower angle (which is obtained by 
translation in the second factor of $SU(2)\times SU(2)$).}
\end{figure}

The geometry of the brane at the critical angle does not have any
non-trivial two-cycle, so there is no F-flux that could stabilise the
brane: the brane will decay. Above the critical angle, the
intersections shown in figure~\ref{fig:genperma} destabilise the
brane. One might expect that the brane dissolves at the intersections
and decays into a brane with an angle $\xi$ below the critical one (see
figure~\ref{fig:genpermd}).
\smallskip

The analysis can easily be extended to the case of arbitrary
levels. The critical angle is then given by
\begin{equation}
\xi^{c}=\frac{\pi}{k'_{1}k'_{2}} \ \ .
\end{equation}
What can we learn about brane charges? By
comparison with ordinary untwisted branes in $SU (2)$, it is
suggestive to take the F-flux as counting the charge. Then the first
brane at angle $\xi_{1} =\pi /\kappa$ carries charge 1, the second
brane charge 2 and so on. The $k^{\text{th}}$ brane sits at the critical angle,
$\xi_{k} =\xi^{c}=\pi \frac{k}{\kappa}$, and it should have charge
zero. This means that the brane charge takes values in the group
$\mathbb{Z}_{k}$ and there are $k-1$ branes without self-intersections
with angles $\xi_{n}=\pi \frac{n}{\kappa}$ where $n=1,\dotsc ,k-1$.
This is precisely the charge group which we expected from K-theory,
confirming that the proposed branes carry the missing K-theory
charges.

% -----------------------------------------------------------------------
% -----------------------------------------------------------------------
\subsection{\label{sc:DBI}Dirac-Born-Infeld analysis}

% -----------------------------------------------------------------------
\subsubsection{Preliminaries}

  In the geometric limit of string theory the dynamics of D-branes is
  known to be described by the Dirac-Born-Infeld (DBI) action
  \cite{Fradkin:1985qd,Abouelsaood:1987gd,Callan:1987bc}. In our situation where the dilaton
  is constant, the action functional may be written\footnote{We assume Neumann
  boundary conditions in the time direction and ignore the latter from now on.}
\begin{equation}
  \label{eq:DBI}
  \mc{S}_\text{DBI}\ \sim\ \int_{\mc{D}}d^py\,\sqrt{\det(\hat{g}+\hat{B}+F)}\ \ .
\end{equation}
  The integral is performed over the $p$-dimensional world volume of the
  D-brane and $\hat{g}$ and $\hat{B}$ are the induced metric and the induced
  B-field, respectively. These two terms depend on the embedding
  of the brane into the given string background. In addition we have a two-form
  field $F$ which lives on the brane. The fields $\hat{B}$ and $F$ are not
  gauge invariant but their sum is and it agrees with the boundary two-form
  $\omega_C$ \eqref{eq:OmegaGen} that we found in the Lagrangian
  setting. The DBI approach should be applicable to our generalised
  permutation branes in the limit of large levels. Since the geometry
  heavily depends on the coprime integers $k'_{1}$ and $k'_{2}$, the
  latter should be kept fixed in that limit. 
\smallskip

  The Dirac-Born-Infeld action allows us to extract important information about
  the properties of D-branes. First and most important it tells us which
  submanifolds are allowed as D-branes. The latter correspond to embeddings
  of the $p$-dimensional world volume into the target space which minimise
  the DBI action. By calculating the fluctuations about such a solution
  we can also gain knowledge about the low-energy excitations of the
  brane. In the case of maximally symmetric branes in compact group manifolds
  this program has been
  carried out in \cite{Bachas:2000ik,Bordalo:2001ec}.
  In this section we want to show that our proposal \eqref{eq:TwConjNeq}
  for the geometry of generalised permutation branes indeed minimises
  the functional \eqref{eq:DBI}, thus proving conformal invariance.
\smallskip

  The variation of the brane embedding leads to the gauge invariant
  equations of motion
\begin{equation}
  \label{eq:EOM}
  \bigl[(\hat{g}+\omega_C)^{-1}\bigr]^{ba}\,\Omega_{ab}^\mu\ =\ 0\ \ ,
\end{equation}
  which have been derived in \cite{Ribault:2003sg}. This
  has to be supplemented by the condition
\begin{equation}
  \label{eq:FEOM}
  \partial_b\Bigl( \sqrt{\det(\hat{g}+\omega_C)}
\bigl[(\hat{g}+\omega_C)^{-1}\bigr]_{\text{antisym}}^{ab}\Bigr)
  \ =\ 0
\end{equation}
  arising from the variation of the F-field. In the last two
  equations we denoted by $X^\mu$ the coordinates of the target space
  and by $Y^a$ the coordinates of the brane whose embedding is given by
  $X^\mu=X^\mu(Y^a)$. While the equations \eqref{eq:FEOM} are
  self-instructional the equations \eqref{eq:EOM} require a bit of explanation.
  The main ingredient $\Omega_{ab}^\mu$ is a generalisation of the
  second fundamental form and takes into account
  the background fluxes. We may write
\begin{equation}
  \label{eq:FundForm}
  \Omega_{ab}^\mu
  \ =\ \partial_a\partial_b X^\mu+\Gamma_{\nu\rho}^\mu\partial_a X^\nu\partial_b X^\rho
       -\hat{\Gamma}_{ab}^c\partial_c X^\mu\ \ .
\end{equation}
  The connections which enter the definition of this object
  are a combination of the Levi-Civita connection and the three-form flux,
\begin{align}
  \label{eq:GenConn}
  \Gamma&\ =\ \Gamma(g)-\frac{1}{2}H&
  \Gamma(g)_{\lambda\mu\nu}
  &\ =\ \frac{1}{2}\bigl(\partial_\mu g_{\lambda\nu}+\partial_\nu g_{\lambda\mu}-\partial_\lambda g_{\mu\nu}\bigr)\ \ .
\end{align}
  Hatted objects refer to induced quantities. In order to calculate
  $\hat{\Gamma}$ we use the previous formulas and plug in
  the induced H-field $\hat{H}$ and the Levi-Civita connection for
  the induced metric $\hat{g}$.
\smallskip

% -----------------------------------------------------------------------
\subsubsection{\label{sc:DBIcalc}Calculations for the branes in $SU(2)_{k_1}\times SU(2)_{k_2}$}

  In order to check that the generalised permutation branes satisfy 
  the equations of motion \eqref{eq:EOM} and~\eqref{eq:FEOM} a detailed
  knowledge of the induced metric, the induced H-field and the boundary
  two-form is required. The first two quantities are derived from the
  background fields on $SU(2)_{k_1}\times SU(2)_{k_2}$ which we will
  describe now. A particularly convenient parametrisation of $SU(2)$ is
\begin{equation}
  \label{eq:grepconcrete}
  g\ =\ \mat\cos\psi+i\cos\theta\sin\psi&\sin\psi\sin\theta e^{i\phi}\\
            -\sin\psi\sin\theta e^{-i\phi}&\cos\psi-i\cos\theta\sin\psi\tam\ \ .
\end{equation}
  In these coordinates the metric takes the form
\begin{equation}
  \label{eq:Metric}
  ds^2\ =\ k_1\bigl[d\psi_1^2+\sin^2\psi_1(d\theta_1^2+\sin^2\theta_1\,d\phi_1^2)\bigr]
           +k_2\bigl[d\psi_2^2+\sin^2\psi_2(d\theta_2^2+\sin^2\theta_2\,d\phi_2^2)\bigr]\ \ .
\end{equation}
  A straightforward calculation yields the H-field
\begin{equation}
  \label{eq:H}
  H\ =\ 2k_1\sin^2\psi_1\sin\theta_1\ d\psi_1\wedge d\theta_1\wedge d\phi_1
        +2k_2\sin^2\psi_2\sin\theta_2\ d\psi_2\wedge d\theta_2\wedge d\phi_2\ \ .
\end{equation}
  Locally this can be integrated and gives rise to a
  B-field of the form
\begin{equation}
  B\ =\ k_1\Bigl(\psi_1-\frac{1}{2}\sin(2\psi_1)\Bigr)\,\sin\theta_1\,d\theta_1\wedge d\phi_1
        +k_2\Bigl(\psi_2-\frac{1}{2}\sin(2\psi_2)\Bigr)\,\sin\theta_2\,d\theta_2\wedge d\phi_2\ \ .
\end{equation}
  This expression is valid on the whole space as long as
$\psi_{i}\not= \pi $.
\smallskip

\begin{figure}
\begin{center}
\input{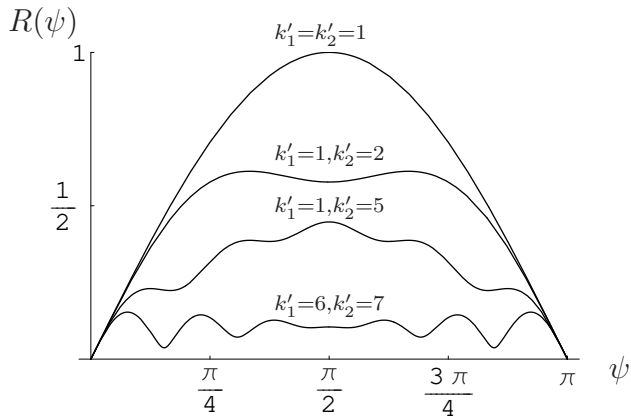}
\end{center}
\caption{\label{fig:defsphere}An illustration of the geometry of the
three-dimensional brane for different values of $k'_{1}$ and $k'_{2}$
as an $S^{2}$-fibration over the interval $[0,\pi]$. The function $R
(\psi)$ measures the radius of the two-sphere sitting over $\psi\in
[0,\pi ]$, $d\hat{s}^{2}\propto d\psi^{2}+R (\psi)^{2}
(d\theta^{2}+\sin^{2}\theta d\phi^{2})$.}
\end{figure}

  The crucial feature of the parametrisation \eqref{eq:grepconcrete}
is that arbitrary powers of $g$ are under explicit control. Namely,
the group element $g^n$, $n\in\Integer$, has the same coordinates
$\theta$ and $\phi$ as $g$ itself. Just the angle $\psi$ has to be
replaced by $n\psi$. In the case of the lowest dimensional generalised
permutation brane this property makes it particularly simple to
determine the induced metric and the induced H-field. Let indeed
$(g_1,g_2)=(g^{k'_2},g^{-k'_1})$ be an element of the brane
\eqref{eq:TwConjNeqId}, parametrised by angles
$\psi_i,\theta_i,\phi_i$ and $\psi,\theta,\phi$, respectively.
According to our previous statement the coordinates are related by
$\psi_1=k'_2\psi$, $\psi_2=-k'_1\psi$, $\theta_1=\theta_2=\theta$ and
$\phi_1=\phi_2=\phi$. The induced metric thus reads
\begin{equation}
  \label{eq:InducedMetric}
  d\hat{s}^2\ =\ \frac{k_1k_2}{k}\bigl(k'_1+k'_2\bigr)\,d\psi^2+\bigl[k_1\sin^2k'_2\psi+k_2\sin^2k'_1\psi\bigr](d\theta^2+\sin^2\theta\,d\phi^2)\ \ .
\end{equation}
  This geometry corresponds to a deformed sphere (see
  figure~\ref{fig:defsphere} for an illustration). The
  induced H-field can easily be determined to be
\begin{equation}
  \label{eq:InducedH}
  \begin{split}
    \hat{H}
    &\ =\ \frac{k_1k_2}{k}\Bigl[\cos(2k'_1\psi)-\cos(2k'_2\psi)\Bigr]\sin\theta\ d\psi\wedge d\theta\wedge d\phi\ \ .
  \end{split}
\end{equation}
  An explicit integration over the brane world
  volume shows that the total flux is zero. For the special choice
  $k_1=k_2$ the metric \eqref{eq:InducedMetric} reduces to that
  of a sphere. Note that at the same time the induced H-field vanishes.
\smallskip

  Last but not least we have to derive the value of the boundary two-form.
  Our general formula \eqref{eq:OmegaGen} looks rather cumbersome. Fortunately
  it can also be determined more directly by integrating the relatively simple
  expression \eqref{eq:InducedH} for the induced H-field in our adapted
  coordinates. Following this route we easily find
\begin{equation}
  \label{eq:Omega}
  \omega_C
  \ =\ \frac{1}{2}\bigl[k_2\sin(2k'_1\psi)-k_1\sin(2k'_2\psi)\bigr]\sin\theta\ d\theta\wedge d\phi
  \ =\ \hat{B}\ \ .
\end{equation}
As indicated this coincides with the induced B-field meaning that
there is no additional F-field on the brane. The sum of the induced
metric and the boundary two-form which enters the DBI action is given
by
\begin{equation}\label{gplusomega}
\hat{g}+\omega_C
 \ =\ \mat\frac{k_1k_2}{k}(k'_1+k'_2)&0&0\\
0&p (\psi )&q (\psi )\sin\theta\\0&-q (\psi )\sin\theta&p (\psi )\sin^2\theta\tam\ \ ,
\end{equation}
where
\[
p (\psi)\ =\ k_1\sin^2k'_2\psi+k_2\sin^2k'_1\psi \quad ,\quad 
q (\psi)\ =\ \frac{1}{2}\bigl[k_2\sin(2k'_1\psi)-k_1\sin(2k'_2\psi)\bigr]\ \ .
\]
Now we can check the F-field equation of
motion~\eqref{eq:FEOM}. The matrix that enters in this equation is
\begin{equation}
\sqrt{\det(\hat{g}+\omega_C)}
\bigl[(\hat{g}+\omega_C)^{-1}\bigr]_{\text{antisym}}^{ab} =
\sqrt{\frac{k_{1}k_{2} (k_{1}'+k_{2}')}{k (p (\psi)^{2}+q
(\psi)^{2})}} q (\psi) \left(\begin{array}{ccc}
0&0&0\\
0&0&-1\\
0&1&0
\end{array} \right)\ .
\end{equation}
As this expression only depends on $\psi$, but all $\psi$-components
of the matrix are zero, the equation~\eqref{eq:FEOM} is
satisfied.

To check the other DBI equation of motion~\eqref{eq:EOM} for our brane
we have to combine all the given data into the generalised
connections. Since the computations are straightforward but not
particularly enlightening, we decided to collect the main intermediate
steps in appendix \ref{sc:DBIDetails}. Here we just state the final
outcome that the equations~\eqref{eq:EOM} are indeed satisfied.  These
results imply that the submanifold \eqref{eq:TwConjNeqId} describes a
brane in the geometric regime of string theory, i.e.\ a conformal
boundary condition of the underlying world sheet theory.  
\smallskip

  So far we just discussed the degenerate case of the lowest
  dimensional brane. The other branes are not $3$-dimensional
  but $5$-dimensional submanifolds of $SU(2)\times SU(2)$. This
  is a major technical complication because in order to
  determine the induced metric and the other quantities which
  enter the DBI action \eqref{eq:DBI} we have to be able to
  evaluate both, the powers
\begin{align}
  (h_1fh_2^{-1})^{k'_2}\qquad\qquad\text{ and }\qquad\qquad
  (h_2fh_1^{-1})^{k'_1}\ \ ,
\end{align}
  at the same time in a closed form. Since we have not been able to
  achieve this goal up to now, we restricted ourselves to a numerical
  check for a choice of low levels and a random selection of points on the
  group manifold. This numerical analysis gave a further confirmation
  for our claim that not only \eqref{eq:TwConjNeqId} but also the
  higher-dimensional geometries \eqref{eq:TwConjNeq} describe
  true string theory branes.

% -----------------------------------------------------------------------
% -----------------------------------------------------------------------
\subsection{Brane tensions}

  The brane tension is related to the g-factor in boundary conformal field
  theory \cite{Affleck:1991tk,Harvey:1999gq}. On the other hand we can determine this
  tension of a $p$-dimensional brane in the DBI formalism by evaluating
  the integral
\begin{equation}\label{tension}
  E\ =\ ( 2\pi )^{-p} \int \sqrt{\det(\hat{g}+\omega_{C})}\ \ .
\end{equation}
  The normalisation corresponds to assigning the value $E=1$ to the
  D0-brane. Via their correspondence to g-factors the DBI energies
  can provide valuable hints for a CFT description of generalised
  permutation branes.
\smallskip

Let us start with the smallest, three-dimensional brane. 
Inserting the expression~\eqref{gplusomega} for $g+\omega_{C}$ into the
tension~\eqref{tension}, we find
\begin{multline}
\mathcal{E}^{k_{1},k_{2}}\ =\ \frac{k^{3/2}}{2\pi^{2}}\sqrt{k'_{1}k'_{2}
(k'_{1}+k'_{2})}\\
\times \int_{0}^{\pi} d\psi \sqrt{(k'_{1}+k'_{2})
[k'_{1}\sin^{2} (k'_{2}\psi)+k'_{2}\sin^{2}
(k'_{1}\psi)]-k'_{1}k'_{2}\sin^{2} (k'_{1}+k'_{2})\psi } \ \ .
\end{multline}
For equal levels this is easily evaluated to be
\begin{equation}
\mathcal{E}^{k,k}\ =\ \frac{k^{3/2}}{\sqrt{2}\pi} \ \ .
\end{equation}
For $k'_{1}=1$ and $k'_{2}=2$, the tension is given by an elliptic integral,
\begin{equation}
\mathcal{E}^{k,2k}\ =\ \frac{k^{3/2}}{\pi^{2}}\sqrt{6} \left(E \left(\frac{1}{4} \right) +E
\left(\frac{8}{9} \right) \right)\ \ ,
\end{equation}
where $E$ is the complete elliptic integral of the second kind.
\smallskip

The computation of the tension of the higher dimensional branes is more
involved. Numerical studies indicate that the energies are given by
\begin{equation}
  E_{n}^{k_{1},k_{2}}\ =\ \frac{\kappa}{\pi}\sin 
  \left( \frac{\pi n}{\kappa}\right)\cdot \mathcal{E}^{k_{1},k_{2}}
\end{equation}
  for the brane with $\xi_{n}=\pi n/\kappa$. The energies of all
  generalised permutation branes may thus be expressed in terms
  of that of the simplest one.
\smallskip

It is interesting to note that the energy dependence on $n$ is simply
given by a sine-factor which is proportional to the modular S-matrix
of $\widehat{su}(2)$ at level $\kappa-2$. This result can be
translated into a prediction for the g-factors that we should expect
in a CFT analysis. The g-factor is directly proportional to the
tension; by using the correct normalisation (which we can get from a
comparison with the untwisted Cardy branes), we obtain
\begin{equation}\label{gfacpred}
g^{k_{1},k_{2}}_{n}\ =\ (k'_{1}k'_{2})^{-1/4}\frac{\sin \frac{\pi
n}{\kappa }}{\sqrt{\sin \frac{\pi}{k_{1}} \sin \frac{\pi}{k_{2}}}}
\,\frac{\sqrt{2}\pi}{k^{3/2}}\, \mathcal{E}^{k_{1},k_{2}} \ \ .
\end{equation}
This prediction, however, has to be taken with a pinch of salt. The DBI
approach neglects higher order curvature corrections and can be trusted
only for large $k$. In the supersymmetric model, one might hope that the
DBI result predicts the exact CFT data as it is the case for the maximally
symmetric branes. Indeed, for equal levels equation~\eqref{gfacpred} reproduces
precisely the known g-factors of ordinary permutation branes,
\begin{equation}
g^{k,k}_{n}\ =\ \frac{S_{n-1\,0}}{S_{00}}\ \ ,
\end{equation}
where $S$ is the modular S-matrix of $\widehat{su}(2)_{k-2}$.

% -----------------------------------------------------------------------
% -----------------------------------------------------------------------
\subsection{Brane spectrum}

  The semi-classical description of generalised permutation branes in
  $SU(2)\times SU(2)$ that we developed in the
  last few sections may be used to extract valuable information about the
  low-energy excitations of open strings which correspond to particle
  wave functions. Since our branes possess a residual $SU(2)$ symmetry we
  expect the spectrum to be organised in representations of $SU(2)$ and
  indeed this is what we will find.

% -----------------------------------------------------------------------
\subsubsection{Derivation of the Laplacian}

  In the presence of non-trivial fluxes the open strings see a deformed
  geometry. In order to determine the Laplacian on the brane we therefore
  have to use the open string metric (see e.g.\ \cite{Seiberg:1999vs})
\begin{equation}
  G\ =\ \hat{g}-\omega_C\hat{g}^{-1}\omega_C\ \ ,
\end{equation}
   which does not only depend on the induced metric $\hat{g}$ on the brane
  as given in \eqref{eq:InducedMetric},
  but also on the boundary two-form $\omega_C$ which has been specified
  in \eqref{eq:Omega}. Plugging in the explicit values for the
  three-dimensional brane we find
\begin{equation}
  \label{eq:OpenMetric}
  G\ =\ \mat c&0&0\\0&a(\psi)&0\\0&0&a(\psi)\sin^2\theta\tam\ \ ,
\end{equation}
  where the function $a(\psi)$ and the constant $c$ are given by
\begin{align}
  \label{eq:Coeffs}
  c&\ =\ k'_1k'_2(k_1+k_2)\\ 
  a(\psi)&\ =\ \frac{4[k_1\sin^2(k'_2\psi)+k_2\sin^2(k'_1\psi)]^2+[k_1\sin^2(2k'_2\psi)-k_2\sin^2(2k'_1\psi)]^2}{4[k_1\sin^2(k'_2\psi)+k_2\sin^2(k'_1\psi)]}\ \ .
\end{align}
  The metric \eqref{eq:OpenMetric} clearly exhibits the spherical
  symmetry we expected from the start. In order to determine the
  spectrum we have to identify the normalisable eigenfunctions
  of the Laplacian
\begin{equation}
  \Delta f
  \ =\ -\frac{1}{\sqrt{G}}\partial_a\bigl[\sqrt{G}G^{ab}\partial_bf\bigr]
  \ =\ -\frac{1}{c}\Bigl[\frac{a'(\psi)}{a(\psi)}\partial_\psi+\partial_\psi^2\Bigr]f
       +\frac{1}{a(\psi)}\Delta_{S^2}f\ \ .
\end{equation}
  Note that the dependence on $\theta$ and $\phi$ could be completely absorbed
  in the Laplacian $\Delta_{S^2}$ on a sphere.
\smallskip

  From the general structure of the previous expression we see that we
  can use the separation ansatz 
\begin{equation}
f_{lm}(\psi,\theta,\phi)=g_{lm}(\psi)Y_{lm}(\theta,\phi)
\end{equation}
  which involves the spherical harmonics $Y_{lm}(\theta,\phi)$. Using the
  eigenvalue equation of the latter with respect to $\Delta_{S^2}$ we arrive at
  the ordinary differential equation
\begin{equation}
  \label{eq:SpectrumODE}
  a'(\psi)g'_{lm}+a(\psi)g''_{lm}-\bigl[l(l+1)c-\lambda ca(\psi)\bigr]g_{lm}\ =\ 0\ \ ,
\end{equation}
  which determines the spectrum of eigenvalues $\lambda$ on our brane. Since
  the function $a(\psi)$ is rather complicated it seems hopeless to solve this
  equation in full generality in terms of known functions. As we will see
  below the solution may be expressed in terms of associated Legendre functions
  in the case of equal levels. For the remaining cases, however, we restrict
  ourselves to a numerical analysis and to an estimate in the regime of large
  eigenvalues.

% -----------------------------------------------------------------------
\subsubsection{The case of equal levels}

  Let us now specialise to the case $k_1=k_2=k$. For these parameters
  eq.\ \eqref{eq:OpenMetric} reduces to
\begin{equation}
  G\ =\ \mat2k&0&0\\0&2k\sin^2(\psi)&0\\0&0&2k\sin^2(\psi)\sin^2\theta\tam\ \ .
\end{equation}
  This is simply the metric for an ordinary $3$-sphere at radius $\sqrt{2k}$.
  Since $S^3\sim SU(2)$ we know that in this case the Laplacian is proportional
  to the quadratic Casimir operator and its spectrum of eigenfunctions
  can be deduced from the Peter-Weyl theorem. More precisely, the algebra of
  functions may be decomposed into irreducible representations $\mc{H}_j$ of
  $SU(2)$ according to
\begin{equation}
  \mc{F}\bigl(S^3\bigr)\ =\ \bigoplus_{j=0}^\infty\,(2j+1)\,\mc{H}_j\ \ .
\end{equation}
  The semi-classical spectrum should be proportional to the quadratic
  Casimirs $j(j+1)$ where each eigenvalue has a
  degeneracy of $(2j+1)^2$. As one may easily check, this prediction coincides
  with the CFT calculation for the $3$-dimensional permutation brane in
  $SU(2)_k\times SU(2)_k$ in the limit of large level $k$ (see
  e.g.\ \cite{Recknagel:2002qq,Alekseev:2002rj}).
\smallskip

  In order to get some intuition how the solutions of
  eq.\ \eqref{eq:SpectrumODE} look like, we would like to construct
  them explicitly and confirm the previous predictions. Using the ansatz
\begin{equation}
  g(\psi)\ =\ (\sin\psi)^{-\frac{1}{2}}h\bigl(\cos(\psi)\bigr)
\end{equation}
  and substituting $z=\cos(\psi)$,
  the original differential equation assumes the form
\begin{equation}
  \label{eq:Legendre}
  (1-z^2)h''-2zh'
  +\Bigl\{\nu(\nu+1)-\frac{\mu^2}{1-z^2}\Bigr\}h\ =\ 0
\end{equation}
  of Legendre's differential equation with parameters given
  by
\begin{align}
  \mu&\ =\ -\Bigl(l+\frac{1}{2}\Bigr)&
  \nu&\ =\ -\frac{1}{2}+\sqrt{1+2k\lambda}\ \ .
\end{align}
  The solutions are associated Legendre functions $P_\nu^\mu(z)$
  and the requirement of regularity of our solutions at $z=\pm1$
  leads to the constraint 
\begin{equation}
  \label{quantcond}
  \mu+\nu\ =\ n\in\Natural_0\ \ .
\end{equation}
  Solving for the eigenvalue $\lambda$ we obtain
\begin{equation}
  \lambda\ =\ \frac{(n+l)(n+l+2)}{2k}\ =\ \frac{2j(j+1)}{k}\ \ .
\end{equation}
  In the last step we identified $n+l$ with twice the spin $j$. Let us
  finally determine the number of solutions with given eigenvalue
  $\lambda$. For fixed $l$ we have a degeneracy $2l+1$ coming from the
  spherical harmonics. Altogether we thus have a multiplicity of
\begin{equation}
  \sum_{l=0}^{2j}(2l+1)\ =\ (2j+1)^2\ \ .
\end{equation}
  This precisely confirms our expectations.

% -----------------------------------------------------------------------
\subsubsection{The general case and solutions for large eigenvalues $\lambda$}

  For a detailed discussion of the differential equation
  \eqref{eq:SpectrumODE} the ansatz
\begin{equation}
  g(\psi)\ =\ u(\psi)/\sqrt{a(\psi)}
\end{equation}
  turns out to be very useful since it eliminates the first order
  term. The resulting equation for $u(\psi)$ is given by
\begin{align}
  \label{eq:NewSpectrumODE}
  u''-f(\psi)u&\ =\ 0&
  &\text{ with }&
  f(\psi)&\ =\ \frac{2a''a-a^{\prime2}}{4a^2}+l(l+1)\frac{c}{a}-\lambda c\ \ .
\end{align}
  From the definition \eqref{eq:Coeffs} we infer that
  $a(\psi)$ is periodic with period $\pi$. Note also that the function $f(\psi)$ may
  be rewritten as a rational function in the variable $z=\cos(\psi)$,
  but the powers involved in the numerator and in the denominator
  will depend in a non-trivial way on the levels
  $k_1$ and $k_2$ and can be rather large. In fact the maximal power
  is not bounded when considered as a function of $k_1$ and $k_2$.
\smallskip

  As a consequence we will not try to solve equation
  \eqref{eq:NewSpectrumODE} in closed form in full generality. In a
  case by case study one might hope to find all the solutions
  explicitly for certain values of the levels. Here, however, we
  will focus our attention to the case of large eigenvalues $\lambda$.
  Under this assumption $\lambda$ will dominate the function $f(\psi)$
  except when the latter becomes singular which happens precisely at the
  boundary $\psi=0,\pi$. Close to $\psi=0$ the function $a(\psi)$
  possesses an expansion
\begin{align}
  a(\psi)&\ =\ \gamma^{-1}\,\psi^2+O(\psi^4)&
  &\text{ with }&
  \gamma^{-1}&\ =\ \frac{k'_1k'_2(5k_1^2-6k_1k_2+5k_2^2)}{k_{1}+k_{2}}\ \ .
\end{align}
  After calculating the derivatives one recognises that the first
  term in $f(\psi)$ is subleading at $\psi=0$. This observation
  motivates us to consider the differential equation
\begin{align}
  \label{eq:RedSpectrumODE}
  u''-f_s(\psi)u&\ =\ 0&
  &\text{ with }&
  f_s(\psi)&\ =\ \frac{l(l+1)\gamma c}{\sin^2(\psi)}-\lambda c\ \ ,
\end{align}
  instead of the original one. We replaced $f(\psi)$ by its singular
part $f_s(\psi)$ which differs from $f (\psi )$ only by a regular
function independent of $\lambda$. Dropping this regular part will
change the eigenvalue $\lambda$ only by an amount which does not grow
with $\lambda$. 
The new differential equation can be solved similarly to
the case at equal levels by transforming it into Legendre's
differential equation \eqref{eq:Legendre}. Here the parameters are
\begin{align}
  \mu&\ =\ -\sqrt{l (l+1)\gamma c+\frac{1}{4}}&
  \nu&\ =\ -\frac{1}{2}+ \sqrt{\lambda c}\ \ .
\end{align}
With the quantisation condition~\eqref{quantcond} we find for the
eigenvalues $\lambda$ of the reduced equation~\eqref{eq:RedSpectrumODE} 
\begin{equation}
\lambda_{n}^{\text{approx}} c\ =\ \bigg(n+\frac{1}{2}+\sqrt{l (l+1)\gamma
c+\frac{1}{4}} \bigg)^{2} \ \ .
\end{equation}
The true eigenvalues $\lambda_{n}$ differ from these by a shift which
becomes independent of $n$,
\begin{equation}
\lambda_{n} c\ =\ d (k'_{1},k'_{2},l)+\bigg(n+\frac{1}{2}+\sqrt{l (l+1)\gamma
c+\frac{1}{4}} \bigg)^{2} + \ (\text{terms vanishing for $n\to \infty$}) \ \ .
\end{equation}
Comparing with the numerical results which are discussed below,
this result describes the spectrum very well for large eigenvalues $\lambda$. 

% -----------------------------------------------------------------------
\subsubsection{Numerical analysis}

  If the levels $k_{1}$ and $k_{2}$ are different, we could just identify
  the asymptotic form of the spectrum analytically. Here, we would like
  to determine the energy of the low lying excitations accurately by
  numerical methods. This provides us with an expectation for the lowest
  conformal weights of boundary fields in a CFT formulation. Furthermore,
  we can check whether there are degeneracies of eigenvalues which would
  signal an enhanced symmetry.
\smallskip

\begin{figure}
  \begin{center}
    \includegraphics[width=8cm]{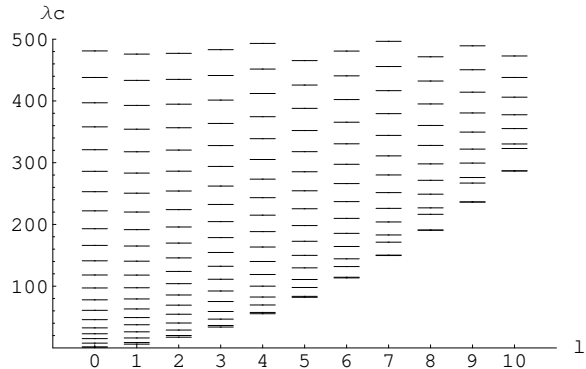}
    \caption{\label{fig:Spectrum}The spectrum of the $3$-dimensional
    generalised permutation brane on $SU(2)_1\times SU(2)_2$.}
  \end{center}
\end{figure}
In the following we compute the spectrum numerically for the
simplest non-trivial case of $k_1=1$ and $k_2=2$. As we mentioned
before the differential equation \eqref{eq:NewSpectrumODE} is periodic
with period $\pi$ and invariant under $\psi\mapsto\pi -\psi$. The
solutions thus split into two classes, those which
are even and those which are odd under $\psi\mapsto\pi -\psi$.  We
determined numerically for which values of $\lambda$ the function with
boundary values
\begin{align}
  i)&\quad u(\pi/2)=1,\ u'(\pi/2)=0&
  ii)&\quad u(\pi/2)=0,\ u'(\pi/2)=1
\end{align}
  can be extended to the boundary $\psi=\pi$ such that $u(\pi)=0$.
  For values of $\lambda$ which are not part of the spectrum the
  solution is singular and $u(\psi)$ will blow up. For $\lambda$
  sufficiently close to an eigenvalue, the solution, however, will
  be sufficiently well-behaved close to $\psi=\pi$. Our findings
  are summarised in table \ref{tb:Spectrum} and plotted in figure
  \ref{fig:Spectrum}. Our investigation shows that the large degeneracy
  of the eigenvalues
  that exists for equal levels is partially lifted. In the generic case
  only the $SU(2)$-degeneracy associated with the spherical harmonics survives.

\begin{table}
  \begin{center}
    \begin{tabular}{|c|cccccccccc|}\hline
      Quantum number $l$ & $0$ & $0$ & $1$ & $0$ & $1$ & $0$ & $1$ & $2$ & $2$ & 0 \\\hline
      Parity (even/odd) & e & o & e & e & o & o & e & e & o & e \\\hline
      Energy $c\lambda$ & $0$ & $1.78$ & $6.14$ & $7.75$ & $8.70$ & $15.07$ & $16.14$ & $17.38$ & $20.14$ & $22.92$ \\\hline
      Multiplicity & $1$ & $1$ & $3$ & $1$ & $3$ & $1$ & $3$ & $5$ & $5$ & $1$ \\\hline
    \end{tabular}
    \caption{\label{tb:Spectrum}Low lying excitations of the $3$-dimensional
    generalised permutation brane on $SU(2)_1\times SU(2)_2$.}
  \end{center}
\end{table}

% -----------------------------------------------------------------------
% -----------------------------------------------------------------------
% -----------------------------------------------------------------------
\section{\label{sc:HighRank}Generalisation to higher rank groups}

  In the last section of this paper we would like to indicate how
  the previous ideas may be generalised to the product $G\times G$ of
  simple groups of higher rank. The presentation will closely follow the
  exposition in appendix~A of \cite{Maldacena:2001xj}. We restrict
  ourselves to the case of the lowest dimensional generalised permutation
  brane.

% -----------------------------------------------------------------------
% -----------------------------------------------------------------------
\subsection{A useful parametrisation of simple groups}

  Let $G$ be a compact simple simply-connected Lie group. As we saw for the
  simplest generalised permutation brane in $SU(2)\times SU(2)$, it is
  absolutely crucial to find a coordinate system in which an arbitrary power of a group element
  $g\in G$ can easily be determined if we want to perform explicit calculations.
  We follow \cite{Maldacena:2001xj} and use the parametrisation
\begin{equation}
  \label{eq:grep}
  g\ =\ h^{-1}(\theta)t(\chi)h(\theta)\ \ ,
\end{equation}
  where $h\in G$ and $t$ is an element of the Cartan torus $T$.
  There is some obvious redundancy under the replacement $h\mapsto fh$ 
  with $f\in T$, such that $h$ in fact takes values in the coset $G/T$.
  Note that powers of $g$ are easily determined to be $g^n=h^{-1}t^nh$.
  In this sense the parametrisation \eqref{eq:grep} mimics the behaviour
  of the coordinates which we used on $SU(2)\times SU(2)$ in the previous
  section. Let us emphasise that in contrast to \cite{Maldacena:2001xj} we
  are not assuming $t$ to be constant.
\smallskip

  Our aim is to express the metric and the Wess-Zumino form in
  coordinates related to the decomposition \eqref{eq:grep} above.
  Let us introduce a Cartan-Weyl basis of $\g$ consisting of Cartan elements
  $H_i$ and root generators $E^\alpha$. They satisfy the commutation
  relations
\begin{align}
  \label{eq:CommRel}
  [H_i,E^\alpha]&\ =\ \alpha^i\,E^\alpha&
  [E^\alpha,E^{-\alpha}]&\ =\ \alpha^i\,H_i
\end{align}
  and others which will not be important in the sequel but can be
  found in any text book or in \cite{Maldacena:2001xj}. The operators
  are assumed to be orthonormal,
\begin{align}
  \tr(H_iH_j)&\ =\ \delta_{ij}&
  \tr(E^\alpha E^\beta)&\ =\ \delta_{\alpha,-\beta}\ \ ,
\end{align}
  where `$\tr$' denotes a suitably normalised trace. We will assume
  $t$ to be of the form $t=e^{i(\chi,H)}$.
\smallskip

  For the purpose of explicit
  calculations it is convenient to introduce the one-forms
\begin{align}
  \label{eq:DefRel}
  t^{-1}dt&\ =\ dtt^{-1}\ =\ i(d\chi,H)&
  &\text{ and }&
  dhh^{-1}&\ =\ \sum_{\alpha>0}\Bigl[\theta^\alpha E^\alpha-\bar{\theta}^\alpha E^{-\alpha}\Bigr]+i(\zeta,H)\ \ .
\end{align}
  For group elements $g$ close to the identity we will use $\chi$ and the
  complex numbers $\theta^\alpha$ as coordinates. The coordinate $\zeta$ is
  an auxiliary variable which will drop out of every physical expression
  in accordance with the gauge symmetry present in \eqref{eq:grep}.
  In terms of the coordinates just introduced the (normalised) metric
  $ds^2=-\frac{k}{2}\tr\bigl[g^{-1}dg\otimes g^{-1}dg\bigr]$ admits the
  simple form
\begin{equation}
  ds^2\ =\ 2k\sum_{\alpha>0}\,\sin^2\frac{1}{2}(\chi,\alpha)\,
           \bigl[\theta^\alpha\otimes\bar{\theta}^\alpha+\bar{\theta}^\alpha\otimes\theta^\alpha\bigr]
           +\frac{k}{2}\sum_md\chi^m\otimes d\chi^m\ \ .
\end{equation}
  If we had assumed $\chi$ to be constant, we would exactly recover
  the result of \cite{Maldacena:2001xj}. During the calculation we made use
  of the commutativity of elements in the Cartan subalgebra and we
  employed the formula
\begin{equation}
  \label{eq:AdjointFormula}
  \Ad_t(dhh^{-1})
  \ =\ e^{i(\chi,\alpha)}\,\theta^\alpha E^\alpha-e^{-i(\chi,\alpha)}\,\bar{\theta}^\alpha E^{-\alpha}+i(\zeta,H)\ \ ,
\end{equation}
  which follows from the commutation relations \eqref{eq:CommRel}.
  This relation will be crucial in the following.
\smallskip

  The same techniques can be used to calculate the Wess-Zumino
  form. With the help of the Polyakov-Wiegmann identity \eqref{eq:PW}
  we find
\begin{equation}
  \begin{split}
    \omega^{\text{WZ}}(g)\bigr|_{g=h^{-1}th}
    &\ =\ -\frac{k}{2}\,d\,\tr\bigl[\Ad_t(dhh^{-1})\wedge dhh^{-1}+2t^{-1}dt\wedge dhh^{-1}\bigr]\ \ .
  \end{split}
\end{equation}
  The term $\omega^{\text{WZ}}(t)$ drops out because $t$ is an element of the
  Cartan torus. Using eq.\ \eqref{eq:AdjointFormula} we may further
  simplify the previous expression. This manipulation yields\footnote{It is not
  entirely surprising that this expression depends on the unphysical parameter
  $\zeta$. In fact, we could never expect the Wess-Zumino form to be exact.
  The occurrence of $\zeta$ is a remnant of this observation. The dependence on $\zeta$
  should disappear if we expand the expression as a genuine three-form using the
  Maurer-Cartan equations.}
\begin{equation}
  \label{eq:GenWZ}
  \omega^{\text{WZ}}(g)\bigr|_{g=h^{-1}th}
  \ =\ k\,d\,\Bigl[i\sum_{\alpha>0}\sin(\chi,\alpha)\,\theta^\alpha\wedge\bar{\theta}^\alpha+(d\chi,\zeta)\Bigr]\ \ .
\end{equation}
  In the case of a constant $\chi$ one can immediately read off the boundary
  two-form for untwisted branes on the group manifold $G$. In this paper,
  however, we are interested in generalised permutation branes in the product
  $G\times G$ and thus there is still some work to do.

% -----------------------------------------------------------------------
% -----------------------------------------------------------------------
\subsection{Application to the product case}

  Let us now consider the product group $G\times G$ at levels $k_1$
and $k_2$.  If we choose real coordinates $\chi_i$ and complex
coordinates $\theta_i^\alpha$ for the elements $g_i$ of the two
individual factors as in the previous section, then the metric is
given by
\begin{equation}
  ds^2\ =\ \sum_{j=1,2}k_j\Biggl\{2\sum_{\alpha>0}\sin^2\frac{1}{2}(\chi_j,\alpha)\,
           \bigl[\theta_j^\alpha\otimes\bar{\theta}_j^\alpha+\bar{\theta}_j^\alpha\otimes\theta_j^\alpha\bigr]
           +\frac{1}{2}\,\sum_{m}\,d\chi_j^m\otimes d\chi_j^m\Biggr\} \ \ .
\end{equation}
  Since we have chosen our parametrisation such that $g^n=h^{-1}t^nh$,
  we can easily determine the induced metric on the generalised permutation
  brane \eqref{eq:TwConjNeqId}. We just have to set $\chi_1=k'_2\chi$ and
  $\chi_2=-k'_1\chi$ while
  $\theta_1^\alpha=\theta_2^\alpha=\theta^\alpha$.\footnote{We should also add
  $\zeta_1=\zeta_2=\zeta$ for completeness.}
  The induced metric for our simplest branes is thus
\begin{align}
  d\hat{s}^2 &\ =\  2\sum_{\alpha>0}\,\biggl[k_1\,\sin^2\frac{k'_2}{2}(\chi,\alpha)+k_2\,\sin^2\frac{k'_1}{2}(\chi,\alpha)\biggr]\,
           \bigl[\theta^\alpha\otimes\bar{\theta}^\alpha+\bar{\theta}^\alpha\otimes\theta^\alpha\bigr]\nonumber\\
           &\qquad +\frac{k_1k_2}{2k}\bigl(k'_1+k'_2\bigr)\sum_m\,d\chi^m\otimes d\chi^m\ \ .
\end{align}
  The Wess-Zumino form may be calculated by adding up the contributions
  \eqref{eq:GenWZ} for the two individual factors,
\begin{equation}
  \omega^{\text{WZ}}(g_1,g_2)
  \ =\ \sum_{j=1,2}\,k_j\,d\,\biggl[i\sum_{\alpha>0}\sin(\chi_j,\alpha)\,\theta_j^\alpha\wedge\bar{\theta}_j^\alpha+(d\chi_j,\zeta_j)\biggr]\ \ .
\end{equation}
  If we restrict this expression to the brane,
  the terms with the unphysical parameters $\zeta_i$ cancel out and
  one can immediately read off the boundary two-form
\begin{equation}
  \omega_C
  \ =\ i\bigl[k_1\sin k'_2(\chi,\alpha)-k_2\,\sin k'_1(\chi,\alpha)\bigr]\,\theta^\alpha\wedge\bar{\theta}^\alpha\ \ .
\end{equation}
  The relevant combination of metric and boundary two-form is thus given by
\begin{align}
  \hat{g}+\omega_C
  &\ =\ \frac{k_1k_2}{2k}\bigl(k'_1+k'_2\bigr)\sum_md\chi^m\otimes d\chi^m
     \nonumber   \\
& \qquad +i\sum_{\alpha>0}\Bigl[k_1\sin k'_2(\chi,\alpha)-k_2\,\sin k'_1(\chi,\alpha)\Bigr]    
            \bigl(\theta^\alpha\otimes\bar{\theta}^\alpha-\bar{\theta}^\alpha\otimes\theta^\alpha\bigr)\nonumber\\
  &\qquad+2\sum_{\alpha>0}\biggl[k_1\,\sin^2\frac{k'_2}{2}(\chi,\alpha)+k_2\,\sin^2\frac{k'_1}{2}(\chi,\alpha)\biggr]\,
            \bigl(\theta^\alpha\otimes\bar{\theta}^\alpha+\bar{\theta}^\alpha\otimes\theta^\alpha\bigr)\ \ .
\end{align}
  This may also be written as
\begin{align}
\hat{g}+\omega_C
    &\ =\ \frac{k_1k_2}{2k}\bigl(k'_1+k'_2\bigr)\sum_md\chi^m\otimes d\chi^m
        \nonumber  \\
&\qquad +\sum_{\alpha>0}\Bigl[k_1\,\bigl(1-e^{-ik'_2(\chi,\alpha)}\bigr)+k_2\,\bigl(1-e^{ik'_1(\chi,\alpha)}\bigr)\Bigr]\,\theta^\alpha\otimes\bar{\theta}^\alpha\nonumber\\[0mm]
    &\qquad+\sum_{\alpha>0}\Bigl[k_1\,\bigl(1-e^{ik'_2(\chi,\alpha)}\bigr)+k_2\,\bigl(1-e^{-ik'_1(\chi,\alpha)}\bigr)\Bigr]\,\bar{\theta}^\alpha\otimes\theta^\alpha\ \ .
 \end{align}
  The structure of this matrix is very simple,
\begin{equation}
  \hat{g}+\omega_C\ =\ \mat A&0&0\\0&0&B\\0&\bar{B}&0\tam\ \ ,
\end{equation}
  with diagonal matrices $A$ and $B$, and the determinant can easily
  be calculated,
\begin{multline*}
  \det(\hat{g}+\omega_C)
  \ =\ 2^{|\Delta_+|}\biggl[\frac{k_1k_2}{k}\bigl(k'_1+k'_2\bigr)\biggr]^r 
\prod_{\alpha>0}\Bigl\{(k_1+k_2)^2+k_1k_2\bigl[\cos(k'_1+k'_2)(\chi,\alpha)-1\bigr]\\
-(k_1+k_2)\bigl[k_1\cos k'_2(\chi,\alpha)+k_2\cos k'_1(\chi,\alpha)\bigr]
\Bigr\}\ \ .
\end{multline*}
  This expressions may be used to partially check the DBI equations of
motion. We refrain from calculating the generalised second fundamental
form \eqref{eq:FundForm} which is needed to verify the eqs.\
\eqref{eq:EOM}. The validity of the F-field equations \eqref{eq:FEOM},
however, follows immediately from the fact that the only dependence of
$\hat{g}+\omega_C$ is on $\chi$ and that the corresponding entries in
its inverse cancel out due to the antisymmetrisation.\footnote{This
argument is not entirely correct because one has to check that the
one-forms $\theta$ provide good coordinates. A more careful analysis
has by now been accomplished in~\cite{Fredenhagen:2009hx}.}

% -----------------------------------------------------------------------
% -----------------------------------------------------------------------
% -----------------------------------------------------------------------
\section{Summary and outlook}

  In this paper we presented evidence for the existence of new branes
  in the product $G\times G$ of Lie groups at different levels. They
  generalise the conventional permutation branes which exist for equal
  levels, but in contrast to the latter they are not maximally
  symmetric. For the string background $SU(2)_{k_1}\times SU(2)_{k_2}$
  we could show explicitly that the proposed branes are solutions of
  Dirac-Born-Infeld theory. Moreover, their tensions revealed a
  surprising link to quantities connected with a single $SU (2)$ model
  at level $\kappa=\lcm(k_1,k_2)$. We also discussed the spectrum of
  excitations of the open string and found that the group theoretical
  degeneracy is partially lifted compared to the case of equal levels.
  Last but not least, the geometry of the branes suggests a very natural
  and complete explanation of the K-theory charges for this particular
  product group.  We hope that the semi-classical data and observations
  we compiled here will facilitate a treatment of these branes in the
  framework of exact CFT in the near future.  
\medskip

  The basic construction we presented in this paper calls for
generalisations in many different directions. First of all it is
straightforward to combine the twist~\eqref{eq:Twist} with
automorphisms which act on the single factors of the product
group. The necessary modification of our formulas is obvious.
As a consequence every pair of automorphisms of $G$ yields not only the
usual factorising branes, but in addition the same number of new
permutation branes. In the case of $SU(3)\times SU(3)$, these branes
already account for all K-theory charges. For a general simple group $G$,
the situation is less obvious. If we knew how to construct all branes
in $G$ required by K-theory, then we would easily obtain all charge
carrying factorising branes in the product group $G\times G$. They,
however, could only account for half of the charges.
It is thus likely that to every
construction of factorising branes there is a corresponding
``permuted'' construction which contributes the remaining charges.
Similar considerations can be applied to
product groups consisting of more than two identical simple group
factors. In that case the permutation group allows cycles of higher
order which will lead to new classes of generalised permutation
branes.
\smallskip

  It is evident that our work just provides a glimpse onto phenomena
in a rich but vastly unexplored landscape of models. In particular,
product groups $G\times G$ in general cannot be part of a consistent
string theory background -- except for $SU(2)\times SU(2)$, Abelian
groups and some low level examples -- already for dimensional
reasons. Instead what we are truly interested in are coset theories
and products thereof.  We are convinced that our generalised
permutation branes have a direct analogue in coset spaces. In fact it
is already known for a long time that a large class of cosets arises
at the boundary of moduli space of current-current deformations of
WZNW models \cite{Giveon:1993ph}.  Such deformations can also be
performed in the presence of a brane and it was shown using
semi-classical methods \cite{Forste:2001gn} that the maximally
symmetric branes in $SU(2)$ are deformed into the $A$- and $B$-type
branes of the parafermions $\text{PF}=SU(2)/U(1)$
\cite{Maldacena:2001ky}.  
\smallskip

  We expect that a similar reasoning is applicable if one starts with
one of our generalised permutation branes in $SU(2)_{k_1}\times
SU(2)_{k_2}$ and deforms the background geometry. At the endpoint of
the deformation one would be left with a presumably non-factorising
brane in the product coset
$\text{PF}_{k_1}\times\text{PF}_{k_2}$.\footnote{Using the
same idea, one could probably even have ``permutation branes'' in
product CFTs such as $SU(2)_{k_1}\times\text{PF}_{k_2}$ where one
factor is a WZNW model while the other one is a coset.} At least we
know that non-factorising permutation branes in the product of
parafermions exist for $k_1=k_2$ and there is no reason to believe
that this should be different if the levels are distinct. A natural
proposal for the lowest dimensional generalised permutation brane in a
coset $G_{k_1}/H\times G_{k_2}/H$ appears to be
\begin{equation}
  \mc{D}_\tau(e,e)
  \ =\ \bigl\{\bigl((gh)^{k'_2},(hg)^{-k'_1}\bigr)\,\bigl|\,g\in G,\,h\in H\bigr\}
  \ \subset\ G\times G
\end{equation}
with $k'_i\ =\ k_i/\gcd(k_1,k_2)$.
  This set is indeed invariant under the adjoint action of
  $H\times H$ and defines therefore a candidate for a brane on the coset. 
  Preliminary
  DBI calculations, however, seem to indicate that this geometry is
  not correct, calling for a more elaborate proposal.
\smallskip

  An extension of our results to cosets is particularly desirable in
view of recent developments in understanding branes in products of
$N=2$ minimal models which, as a coset, are rather similar to the
parafermions. Due to the presence of $N=2$ supersymmetry one can
perform a topological twist and analyse the topological subsector of
the original theory. This simplifies many of the calculations. In
particular, the classification of B-branes can be reduced to the
purely algebraic problem of the classification of matrix
factorisations of the superpotential (this goes back to an unpublished
idea of Kontsevich).  In the corresponding investigations for products
of minimal models a special class of branes has been discovered which
just emerges if the levels involved have common divisors
\cite{Brunner:2005fv}. Moreover, these branes provide an important
contribution to the lattice of K-theory charges~\cite{Caviezel:2005th}.  In
this respect and also in the concrete expressions for the matrix
factorisations they bear a close resemblance to the generalised
permutation branes presented here.  
\smallskip

  The investigation of non-trivial branes in product CFTs has
  interesting applications. In string theory products of
  $N=2$ minimal models arise naturally in Gepner models. 
  Also in statistical physics boundary conditions in products of CFTs 
  play a distinguished role since they may be mapped
  to defect lines between the individual constituents using
  the folding trick \cite{Oshikawa:1997dj}. Up to now, however,
  the classification of conformal defect lines is stuck at a
  rather preliminary stage, see e.g.\ \cite{Quella:2002ct} and
  references therein.
\smallskip

  Summarising the last few paragraphs, there is by now an
  overwhelming evidence for the existence and importance of
  generalised permutation branes from several directions. What
  we still lack is an exact CFT prescription or even an idea what the
  precise infinite dimensional symmetry could be that is preserved
  by the branes. At the moment there is not much we can say
  about these issues but we would at least like to indicate
  where we see the best chances of making progress.
\smallskip

  The most promising candidate for progress on the CFT side seem
  to be the aforementioned products of $N=2$ minimal models. For these
  theories one might
  expect to be able to combine the rather complementary information
  from the topological sector and the semi-classical regime into
  concrete guidelines for identifying the symmetry preserved and
  the construction of boundary states. On the other hand, the
  superconformal symmetry with its severe restrictions might also
  assist in mastering this task. 
\smallskip

  Another interesting idea is to consider generalised
  permutation branes in the product of ordinary minimal models.
  It is known that each minimal model
  $\mc{M}_m$ possesses a relevant integrable perturbation which
  connects it to the minimal model $\mc{M}_{m-1}$ in the infrared.
  In the simplest case of the tricritical Ising model $\mc{M}_4$,
  the flow to the Ising model $\mc{M}_3$ has been analysed in
  detail in \cite{Pearce:2003km}, even in the presence of a
  boundary. One of the main results was an explicit map between
  initial and final boundary conditions along the flow. We
  hope that also the deformation of ordinary permutation
  branes in $\mc{M}_4\times\mc{M}_4$ can be analysed in the same
  spirit if we perturb the model towards $\mc{M}_4\times\mc{M}_3$.
  The major complication compared to the case of a single
  minimal model is that the final brane will probably not
  preserve the whole symmetry $\mc{M}_4\times\mc{M}_3$.
\smallskip

  Let us finally comment on one of the big unsolved problems
  in CFT: the full classification
  of (super)conformal boundary conditions for a given background.
  Only partial results in this direction are available up to now
  \cite{Gaberdiel:2001zq,Janik:2001hb,Cappelli:2002wq,Quella:2002ct,Gaberdiel:2004nv}.
  A detailed knowledge about the CFT construction of generalised
  permutation branes in product groups and cosets would certainly
  be a major step forward. In this paper it turned
  out to be extremely fruitful to accept the guidance of
  geometric and K-theoretic arguments and probably this will
  also be the case for other branes of particular physical
  significance. Eventually, new insights into the
  classification of branes or into the brane charges in certain
  backgrounds could be gained by combining the nested coset
  construction of branes \cite{Quella:2002ct,Quella:2002ns}
  with generalised permutations in common subfactors.
\smallskip

  We hope to return to some of the open issues in future
  publications.

\subsubsection*{Acknowledgements} 
We are very grateful to Anton Alekseev for initial collaboration on this
project. We also appreciate discussions with Costas Bachas, Volker
Braun, Matthias Gaberdiel, Sylvain Ribault, Volker Schomerus and
Gerard Watts.  This work was partially supported by the EU Research
Training Network grants ``Euclid'', contract number
HPRN-CT-2002-00325, ``Superstring Theory", contract number
MRTN-CT-2004-512194, and ``ForcesUniverse'', contract number
MRTN-CT-2004-005104. The work of SF is supported by the Max Planck
Institute for Gravitational Physics and the Max Planck Society. The
research of TQ is financed by a PPARC postdoctoral fellowship under
reference PPA/P/S/2002/00370. He also receives partial support from
the PPARC rolling grant PPA/G/O/2002/00475. This work was initiated
during a stay of SF and TQ at the Section de Math{\'e}matiques, Universit{\'e}
de Gen{\`e}ve, which was enabled by the Swiss National Science Foundation.

\appendix
% -----------------------------------------------------------------------
% -----------------------------------------------------------------------
% -----------------------------------------------------------------------
\section{\label{app:Ktheory}K-theory for products of groups}
In this section we want to determine the twisted K-theory of $G\times G$. For
this we employ the K{\"u}nneth exact sequence (see e.g.\ \cite{Blackadar:1986}), 
\begin{equation}
0\ \longrightarrow\ {}^{\tau_{1}}K (G)\otimes {}^{\tau_{2}}K
(G)\ \longrightarrow\ {}^{\tau_{1}+\tau_{2}}K (G\times G)\ \longrightarrow\ 
\text{Tor}\bigl( {}^{\tau_{1}}K (G),{}^{\tau_{2}}K (G)\bigr)\ \longrightarrow\ 0 \ \ .
\end{equation}
Here, $\tau_{i}$ are elements of the third integral cohomology group
of $G$; these are the Wess-Zumino forms in the corresponding WZNW models.
The sequence splits (unnaturally) so that we can determine
$^{\tau_{1}+\tau_{2}}K (G\times G)$ by computing the tensor product
and the Tor-part. 

The twisted K-theory of a simple, simply connected Lie group has been
determined
in~\cite{Maldacena:2001xj,Braun:2003rd,Freed:2003qx,Douglas:2004}. It
is given by 
\begin{equation}
^{\tau}K (G)=\big(\mathbb{Z}_{d} \big)^{2^{r-1}}
\end{equation}
where
$r$ is the rank of $G$, and the order $d$ depends on $G$ and $\tau$. The
tensor product of two of these K-groups is
\begin{equation}
  {}^{\tau_{1}}K (G)\otimes{}^{\tau_{2}}K (G)
  \ \cong\ \big(\Integer_{\text{gcd} (d_{1},d_{2})} \big)^{2^{2 (r-1)}}\ \ .
\end{equation}
Tor is the derived functor of the tensor product functor. To compute
it we first have to find a resolution of $^{\tau_{i}}K (G)$, i.e.\ a
free chain complex with zeroth cohomology equal to $^{\tau_{i}}K (G)$,
\begin{equation}
0\ \longrightarrow\ \mathbb{Z}^{2^{r-1}}\ \xrightarrow{\ \cdot
d_{i}\ }\ \underline{\mathbb{Z}}^{2^{r-1}}\longrightarrow 0
\ \ .
\end{equation}
The zeroth position is underlined.
We take the tensor product of the chain complexes,
\begin{equation}
0\ \longrightarrow\ \mathbb{Z}^{2^{2 (r-1)}}\ \xrightarrow{(\cdot d_{1},\cdot
d_{2})}\ \mathbb{Z}^{2^{2 (r-1)}} \oplus
\mathbb{Z}^{2^{2 (r-1)}}\ \xrightarrow{(-d_{2},d_{1})}\ 
\underline{\mathbb{Z}}^{2^{2 (r-1)}}\ \longrightarrow\ 0 \ \ .
\end{equation}
The zeroth cohomology of this complex is 
\begin{equation}
\text{Tor}\bigl(^{\tau_{1}}K (G),^{\tau_{2}}K (G)\bigr)\ \cong \
\big(\mathbb{Z}_{\text{gcd} (d_{1},d_{2})} \big )^{2^{2 (r-1)}}\ \ .
\end{equation}
In total we find
\begin{equation}\label{Kgroup}
^{\tau_{1}+\tau_{2}}K (G\times G)\ \cong \ \big(
\mathbb{Z}_{\text{gcd} (d_{1},d_{2})} \big)^{2^{2r-1}}\ \ .  
\end{equation}
Let us look at the example $G=SU (2)$. For the simple factors, the
twisted K-theory is given by $^{\tau_{i}}K (SU
(2))=\mathbb{Z}_{k_{i}}$, where $k_{i}$ is the level of the
corresponding WZNW model. The twisted K-theory of the product group is then 
\begin{equation}
^{\tau_{1}+\tau_{2}}K\bigl(SU (2)\times SU (2)\bigr)\ \cong \ \mathbb{Z}_{\text{gcd}
(k_{1},k_{2})} \oplus \mathbb{Z}_{\text{gcd}
(k_{1},k_{2})} \ \ ,
\end{equation}
where one of the summands contributes to $K^{0}$ and the other to $K^{1}$.

% -----------------------------------------------------------------------
% -----------------------------------------------------------------------
% -----------------------------------------------------------------------
\section{\label{sc:DBIDetails}Details of the Born-Infeld calculation}

  In this appendix we collect some formulas which are needed to check
the DBI equations of motion~\eqref{eq:EOM} in section
\ref{sc:DBIcalc}. We will present explicit expressions for the
generalised connections \eqref{eq:GenConn} based on the metrics
\eqref{eq:Metric} and \eqref{eq:InducedMetric} as well as on the
H-fields \eqref{eq:H} and \eqref{eq:InducedH}. All expressions in this
appendix are based on the parametrisation \eqref{eq:grepconcrete}.
\smallskip

  Since the embedding of the brane into the target space is expressed
  in terms of a linear relation,
  there is no contribution of the first term to the generalised
  second fundamental form \eqref{eq:FundForm}. The derivatives in the
  second term just contribute constant factors. We write
\begin{equation}
  \label{eq:FundFormShort}
  \Omega_{ab}^\mu
  \ =\ \Gamma_{\nu\rho}^\mu\partial_a X^\nu\partial_b X^\rho
       -\hat{\Gamma}_{ab}^c\partial_c X^\mu
  \ =\ \Omega_{ab}^{(1)\mu}+\Omega_{ab}^{(2)\mu}\ \ .
\end{equation}
  In order to determine the first contribution $\Omega_{ab}^{(1)\mu}$
  we start with the Levi-Civita connection for the individual
  group factors for which we find
\begin{align}
  \Gamma(g)_{\psi_i\theta_i\theta_i}&\ =\ -k_i\sin\psi_i\cos\psi_i&
  \Gamma(g)_{\psi_i\phi_i\phi_i}&\ =\ -k_i\sin\psi_i\cos\psi_i\sin^2\theta_i\nonumber\\[2mm]
  \Gamma(g)_{\theta_i\phi_i\phi_i}&\ =\ -k_i\sin^2\psi_i\sin\theta_i\cos\theta_i&
  \Gamma(g)_{\theta_i\psi_i\theta_i}&\ =\ k_i\sin\psi_i\cos\psi_i\\[2mm]
  \Gamma(g)_{\phi_i\psi_i\phi_i}&\ =\ k_i\sin\psi_i\cos\psi_i\sin^2\theta_i&
  \Gamma(g)_{\phi_i\theta_i\phi_i}&\ =\ k_i\sin^2\psi_i\sin\theta_i\cos\theta_i\ \ .\nonumber
\end{align}
  The remaining non-vanishing entries follow from the symmetry of
  the connection in the last two indices.
  By subtracting the individual H-fields and raising the first index
  we obtain the generalised connection
\begin{align}
  \Gamma^{\psi_i}
  &\ =\ \mat0&0&0\\0&-\sin\psi_i\cos\psi_i&-\sin^2\psi_i\sin\theta_i\\0&\sin^2\psi_i\sin\theta_i&-\sin\psi_i\cos\psi_i\sin^2\theta_i\tam\\[2mm]
\Gamma^{\theta_i}
  &\ =\ \mat0&\cot\psi_i&\sin\theta_i\\\cot\psi_i&0&0\\-\sin\theta_i&0&-\sin\theta_i\cos\theta_i\tam \quad , \quad  
  \Gamma^{\phi_i}
  &\ =\ \mat0&-\csc\theta_i&\cot\psi_i\\\csc\theta_i&0&\cot\theta_i\\\cot\psi_i&\cot\theta_i&0\tam\nonumber \ \ .
\end{align}
  Taking the additional factors into account which come from the derivatives
  in \eqref{eq:FundFormShort} and restricting the coordinates to the brane we easily find
\begin{equation}
  \begin{split}
    \Omega^{(1)\psi_1}
    &\ =\ \mat0&0&0\\0&-\sin(k'_2\psi)\cos(k'_2\psi)&-\sin^2(k'_2\psi)\sin\theta\\0&\sin^2(k'_2\psi)\sin(\theta)&-\sin(k'_2\psi)\cos(k'_2\psi)\sin^2\theta\tam+(k'_2\to-k'_1)\\[2mm]
    \Omega^{(1)\theta_1}
    &\ =\ \mat0&k'_2\cot(k'_2\psi)&k'_2\sin\theta\\k'_2\cot(k'_2\psi)&0&0\\-k'_2\sin\theta&0&-\sin\theta\cos\theta\tam+(k'_2\to-k'_1)\\[2mm]
    \Omega^{(1)\phi_1}
    &\ =\ \mat0&-k'_2\csc\theta&k'_2\cot(k'_2\psi)\\k'_2\csc\theta&0&\cot\theta\\k'_2\cot(k'_2\psi)&\cot\theta&0\tam+(k'_2\to-k'_1)\ \ .
  \end{split}
\end{equation}

  Similar, though slightly more cumbersome, calculations have to be performed
  for the induced quantities. Let us first introduce the abbreviations
\begin{align}
    r(\psi)&\ =\ k_1\sin^2(k'_2\psi)+k_2\sin^2(k'_1\psi)&
    s(\psi)&\ =\ k_2\sin(k'_1\psi)\cot(k'_2\psi)-k_1\cos(k'_1\psi)\\[2mm]
    t(\psi)&\ =\ \sin(2k'_1\psi)+\sin(2k'_2\psi)&
    u(\psi)&\ =\ \sin^2(k'_2\psi)-\sin^2(k'_1\psi)\ \ ,
\end{align}
  which allow a compact representation of our results.
  The matrices of the generalised induced connection where the first
  index has been raised read
\begin{equation}
  \begin{split}
    \hat{\Gamma}^\psi
    &\ =\ -\frac{1}{2(k'_1+k'_2)}\mat0&0&0\\0&t(\psi)&u(\psi)\sin\theta\\0&-u(\psi)\sin\theta&t(\psi)\sin^2\theta\tam\\[2mm]
    \hat{\Gamma}^\theta
    &\ =\ \frac{1}{r(\psi)}\mat0&\frac{k_1k_2}{2k}t(\psi)&\frac{k_1k_2}{2k}u(\psi)\sin\theta\\\frac{k_1k_2}{2k}t(\psi)&0&0\\-\frac{k_1k_2}{2k}u(\psi)\sin\theta&0&-r(\psi)\sin\theta\cos\theta\tam\\[2mm]
    \hat{\Gamma}^\phi
    &\ =\ \frac{1}{r(\psi)\sin\theta}\mat0&-\frac{k_1k_2}{2k}u(\psi)&\frac{k_1k_2}{2k}t(\psi)\sin\theta\\\frac{k_1k_2}{2k}u(\psi)&0&r(\psi)\cos\theta\\\frac{k_1k_2}{2k}t(\psi)\sin\theta&r(\psi)\cos\theta&0\tam\ \ .
  \end{split}
\end{equation}
  Finally we are prepared to state the result for the
  complete generalised second fundamental form 
\begin{align*}
    \Omega^{\psi_1}
    &\ =\ \frac{1}{2(k'_1+k'_2)}\mat0&0&0\\0&k'_2\sin(2k'_1\psi)-k'_1\sin(2k'_2\psi)&-2r(\psi)\sin\theta\\0&2r(\psi)\sin\theta&\bigl[k'_2\sin(2k'_1\psi)-k'_1\sin(2k'_2\psi)\bigr]\sin^2\theta\tam\displaybreak[0]\\[2mm]
    \Omega^{\theta_1}
    &\ =\ \frac{1}{r(\psi)}\mat0&k'_2\sin(k'_1\psi)s(\psi)&k'_2(k_1+k_2)\sin^2(k'_1\psi)\sin\theta\\
k'_2\sin(k'_1\psi)s(\psi)&0&0\\
-k'_2(k_1+k_2)\sin^2(k'_1\psi)\sin\theta&0&0\tam\displaybreak[0]\\[2mm]
    \Omega^{\phi_1}
    &\ =\ \frac{1}{r(\psi)\sin\theta}\mat0&-k'_2(k_1+k_2)\sin^2(k'_1\psi)&k'_2\sin(k'_1\psi)s(\psi)\sin\theta\\
k'_2(k_1+k_2)\sin^2(k'_1\psi)&0&0\\
k'_2\sin(k'_1\psi)s(\psi)\sin\theta&0&0\tam\, .
\end{align*}
  The second set of matrices with superscripts $\psi_2,\theta_2,\phi_2$
  can easily be obtained from the previous ones.
\smallskip

  One finally has to plug all these expressions into the equations
  of motion \eqref{eq:EOM}. For the labels $\mu=\theta_1$ and $\mu=\phi_1$
  they are easily seen to be identically satisfied just because of the
  matrix structure which renders the trace zero. Only for $\mu=\psi_1$
  one has to perform a small calculation to verify the equations of motion.

\def\cprime{$'$} \def\cprime{$'$}
\providecommand{\href}[2]{#2}\begingroup\raggedright\endgroup

\end{document}